\documentstyle[12pt,aaspp4]{article}
\def\lsim{\lower.5ex\hbox{$\; \buildrel < \over \sim \;$}}
\def\gsim{\lower.5ex\hbox{$\; \buildrel > \over \sim \;$}}
\lefthead{}
\righthead{  }

\def\lsim{\lower.5ex\hbox{$\; \buildrel < \over \sim \;$}}
\def\gsim{\lower.5ex\hbox{$\; \buildrel > \over \sim \;$}}

\received{2002 September 8}
\begin{document}

\title{Bypass to Turbulence in Hydrodynamic Accretion Disks: An Eigenvalue Approach }

\author{Banibrata Mukhopadhyay\altaffilmark{1}, 
Niayesh Afshordi\altaffilmark{2}, 
Ramesh Narayan\altaffilmark{3}  }
\altaffiltext{1}{bmukhopa@cfa.harvard.edu}
\altaffiltext{2}{nafshord@cfa.harvard.edu}
\altaffiltext{3}{rnarayan@cfa.harvard.edu} 

\affil{\small Institute for Theory and Computation, Harvard-Smithsonian Center for Astrophysics,
60 Garden Street, MS-51, Cambridge, MA 02138}

\begin{abstract}

Cold accretion disks with temperatures below $\sim 3000$K are
likely to be composed of highly neutral gas. The magnetorotational
instability may cease to operate in such disks, so it is of interest
to consider purely hydrodynamic mechanisms of generating turbulence
and angular momentum transport.  With this motivation, we investigate
the growth of hydrodynamic perturbations in a linear shear flow
sandwiched between two parallel walls.  The unperturbed flow is
similar to plane Couette flow but with a Coriolis force included.
Although there are no exponentially growing eigenmodes in this system,
nevertheless, because of the non-normal nature of the eigenmodes, it
is possible to have a large transient growth in the energy of
perturbations.  For a constant angular momentum disk, we find that the
perturbation with maximum growth is axisymmetric with vertical structure.
The energy grows by more than a factor of 100 for a
Reynolds number $R=300$ and more than a factor of 1000 for $R=1000$.
Turbulence can be easily excited in such a disk, as found in previous
numerical simulations.  For a Keplerian disk, on the other hand,
similar perturbations with vertical structure grow by no more than a factor of 4,
explaining why the same simulations did not find turbulence in this
system.  However, certain other two-dimensional perturbations with no
vertical structure do exhibit modest growth.  For the optimum
two-dimensional perturbation, the energy grows by a factor of
$\sim100$ for $R\sim10^{4.5}$ and by a factor of 1000 for
$R\sim10^6$. Such large Reynolds numbers are hard to achieve in
numerical simulations and so the nonlinear development of these kinds
of perturbations are only beginning to be investigated.  It is
conceivable that these nearly two-dimensional disturbances might lead
to self-sustained three-dimensional turbulence, though this remains to
be demonstrated.  The Reynolds numbers of cold astrophysical disks are
much larger even than $10^6$; therefore, hydrodynamic turbulence may
be possible in disks through transient growth.

\end{abstract}

\keywords {accretion, accretion disk --- hydrodynamics --- turbulence --- instabilities }

\section{Introduction}

The origin of hydrodynamic turbulence is not fully understood.  Many
efforts have been devoted to this problem, beginning with the early
work of Kelvin, Rayleigh and Reynolds toward the end of the nineteenth
century.  However, despite a large number of investigations over the
decades, the key physics is still poorly understood.  One of the
reasons for this is that there is a significant mismatch between the
predictions of linear stability theory and experimental data.  For
example, plane Poiseuille flow is known to become turbulent in the
laboratory at a Reynolds number $R\sim 1000$, whereas theory predicts
that the flow is linearly stable up to $R = 5772$.  An even more
severe discrepancy, one that is of direct interest to astrophysics,
occurs in the case of plane Couette flow.  Laboratory experiments and
numerical simulations show that this flow can become turbulent for $R$
as small as $\sim350$.  However, theoretical analysis shows that the
flow is linearly stable for all $R$ up to infinity.  Such a large
discrepancy indicates that linear stability analysis, based on
eigenspectra, is not the best tool for understanding the onset of
turbulence. In this paper, we pursue a different approach, the
so-called {\it bypass mechanism} to transition, which has been popular
in the fluid mechanics literature (e.g., Farrell 1988; Butler \&
Farrell 1992; Reddy \& Henningson 1993; Trefethen et al. 1993).  We
use this approach to study a possible route to turbulence in
astrophysical disks.

Accretion disks in astrophysics operate by transferring angular
momentum outward by an effective ``viscosity.''  Microscopic molecular
viscosity is completely negligible, so it was recognized more than
three decades ago (Shakura \& Sunyaev 1973; Lynden-Bell \& Pringle
1974) that angular momentum transfer must occur via {\it turbulence}
of some sort.  However, the physical origin of the turbulence was not
identified until the important work of Balbus \& Hawley (1991) who
identified the Magneto-Rotational-Instability (MRI) (originally
discovered by Velikhov 1959; Chandrasekhar 1960) and showed that this
linear instability will operate in the presence of very weak magnetic
fields and will lead to magnetohydrodynamic (MHD) turbulence.  The MRI
is now accepted as the origin of turbulence in most accretion disks.
Hawley, Gammie \& Balbus (1995) showed that the MRI dies out if the
Lorentz force is turned off, and Hawley, Gammie \& Balbus (1996)
showed that the magnetic field dies out when the Coriolis force is
turned off while retaining the Lorentz forces.  In both of these
situations, MHD turbulence is absent.  In two subsequent papers,
Balbus, Hawley \& Stone (1996) and Hawley, Balbus \& Winters (1999)
showed through numerical simulations that, whereas pure hydrodynamic
turbulence is easily triggered in plane Couette flow (which was
already known) and in a constant angular momentum disk, turbulence
does not develop in an unmagnetized Keplerian disk even in the
presence of large initial perturbations.  The authors argued on this
basis that hydrodynamic turbulence cannot contribute to viscosity in
accretion disks.
 
Despite the above important work, there is reason to study
hydrodynamic turbulence in astrophysical disks.  Several accretion
systems are known in which the gas is cold and largely neutral so that
the gas and the magnetic field are poorly coupled. The MRI
then becomes weak or may even cease to operate. In a series of
experiments, Hawley, Gammie \& Balbus (1996) and Fleming, Stone \&
Hawley (2000) showed that for a magnetic Reynolds number, $R_M\sim
10^4$, the magnetic field is depressed and for $R_M=2\times 10^3$ the
magnetic field dies out. Thus, for $R_M\le 10^4$, MHD turbulence and
the associated angular momentum transport switch off.  Examples of
systems in this regime include accretion disks around quiescent
cataclysmic variables (Gammie \& Menou 1998; Menou 2000),
proto-planetary and star-forming disks (Blaes \& Balbus 1994; Gammie
1996; Fromang, Terquem \& Balbus 2002), and the outer regions of disks
in active galactic nuclei (Menou \& Quataert 2001; Goodman 2003). Note that the
magnetic Reynolds number may not be the only relevant parameter that
determines the strength of the MRI; the magnetic Prandtl number may
play a role (Gammie \& Menou 1998), as well as ambipolar diffusion
(Blaes \& Balbus 1994; Menou \& Quataert 2001).  Nevertheless, it
seems reasonable to assume that, in cold astrophysical disks, the MRI
will be sluggish or even absent.  The question then arises: How can
these systems sustain mass transfer in the absence of the MRI?  What
drives their turbulence?

A number of ideas have been discussed in the literature in answer
to this question.  Gammie (1996) argued that the surface layers of
cold protostellar disks would be ionized by cosmic rays and that this
would enable accretion to proceed via the MRI within these layers.
Menou (2000) suggested that angular momentum transport in quiescent
cataclysmic variables could be induced by tidal perturbations from the
binary companion star and showed that this mechanism could explain the
occurrence of longer outburst recurrence times in systems with large
binary mass ratios.  In the case of the outer disks of active galactic
nuclei, Menou \& Quataert (2001) showed that the MRI-stable regions
are nearly always gravitationally unstable, so that the latter might
drive angular momentum transport.  A problem with these ideas is that
each is invoked specifically for a particular class of systems.  While
there is nothing in principle wrong with this, one wonders whether
there may not be some more general mechanism for generating turbulent
transport in MRI-stable astrophysical disks.

Recent laboratory experiments on rotating Couette flow in the narrow
gap limit with linearly stable rotational angular velocity profiles
(similar to Keplerian disks) seem to indicate that turbulence does manage to
develop in such flows (Richard \& Zahn 1999). Longaretti (2002) points
out that the absence of turbulence in the simulations by Balbus, 
Hawley \& Stone (1996) and Hawley, Balbus \& Winters (1999) may be because
of their small effective Reynolds number.
Also, Bech \& Andersson (1997) have shown that turbulence persists 
in numerical simulations of sub-critical rotating flows,  
provided the Reynolds number is very high. Moreover, as already
mentioned, it is well known that linear stability is no guarantee that
a flow (whether rotating or not) will avoid becoming turbulent (for a
detailed discussion see e.g. Swinney \& Gollub 1981; Drazin \& Reid 1983).
In fact, since the celebrated work of Orr (1907), it has been known that
linearly stable flows can exhibit significant {\it transient growth}
in energy for certain initial perturbations.  This fact provides a
possible solution to the problem of explaining hydrodynamic turbulence
in linearly stable systems.  The idea is that the transient growth may
allow perturbations to grow to a non-linear state, after which a
sub-critical transition to turbulence may take place.  This is called
the {\it bypass mechanism} to turbulence.  In the astrophysical
literature, an early application of transient growth may be found in
Goldreich \& Lynden-Bell (1965; see also Goldreich \& Tremaine 1978, 1979).

The physics of transient growth has been discussed by a number of
authors (Farrell 1988; Butler \& Farrell 1992; Reddy \& Henningson
1993; Trefethen et al. 1993), who have shown that the growth results
from the non-normal nature of the associated operator.  
The eigenfunctions of the linearly perturbed system are not
orthogonal but are close to linearly dependent in nature, and as a result
certain linear combinations of the eigenfunctions that are arranged
to nearly cancel initially may develop considerable amplitude at later
time when the degree of cancellation is reduced.  Therefore, even in
the absence of any exponentially growing eigenfunctions, the system is
still able to exhibit transient growth. This idea has been discussed 
in the fluid mechanics literature for a number of years but has only recently been
applied to astrophysical accretion disks. 
Ioannaou \& Kakouris (2001) studied the global
behavior of perturbations in an accretion disk, Chagelishvili et
al. (2003) analysed a local 2-dimensional patch in a disk using a {\it shearing box}
approximation and showed that strong growth is possible, and Tevzadze et al. (2003) showed 
that 3-dimensional perturbations also 
undergo substantial transient growth, provided the vertical scale remains 
of the order of the azimuthal scale. Umurhan \& Regev (2004) studied the non-linear
development of the Chagelishvili et al. (2003) growing mode and Yecko (2004)
studied rotating shearing flows between walls. Recently, Johnson \& Gammie (2005a,b)
studied the evolution of a plane-wave type perturbation 
in thin low-ionization disks. 

The aim of the present study is to further explore the physics of
transient non-normal growth of perturbations in cold accretion disks,
with a view to understanding whether such growth could lead to
hydrodynamic turbulence. 
Our aim is to present the analysis in such a manner that even 
readers from other branch of astrophysics (not familiar with the
conventional fluid dynamical approach) will be able to follow and reproduce the
results.  Along with a companion paper (Afshordi, Mukhopadhyay \& Narayan 2005;
hereafter AMN05), we study both Keplerian and constant angular
momentum disks, as well as plane Couette flow.  Both papers
concentrate on identifying the parameter regimes over which a large
transient growth in energy is possible and studying the nature of the
growing perturbations. While the present paper focuses on an
eigenvalue analysis in Eulerian coordinates of flow between walls,
AMN05 presents a Lagrangian analysis of an infinite shear flow.

The plan of the paper is as follows.  In \S 2, we present our basic
model, beginning with a description of the equilibrium flow, then
discussing the perturbation equations and eigenfunctions, and
introducing the concept of transient energy growth. In \S 3, we
present numerical results obtained using the eigenfunction approach
for a variety of flows: plane Couette flow, constant angular momentum
disk, Keplerian disk.  In \S 4, we explain the physics of the
numerical results by means of analytical and heuristic arguments.
Finally, in \S 5, we discuss the implications of the results.  
In the Appendix, we describe the formalism to compute the transient energy growth.

\section{The Model}

\subsection{Equilibrium Flow}

We consider a small patch of an accreting disk centered on radius
$r_0$ and viewed in a frame orbiting at the angular velocity
$\Omega_0$ of the gas at this radius.  We employ Cartesian coordinates
$(X,Y,Z)$ such that $X=r-r_0$ is in the radial direction,
$Y=r_0(\phi-\phi_0)$ is in the azimuthal direction and $Z$ is in the
vertical direction.

For ease of comparison with classical results in the fluid literature,
we assume that the flow is incompressible, that it extends from $X=-L$
to $+L$, and that there are rigid walls at the two ends with no-slip
boundary conditions.  The flow is unbounded along $Y$ and $Z$. In the
limit $L \ll r_0$, the unperturbed velocity corresponds to a linear
shear of the form
\begin{eqnarray}
\vec{V}=(0,-\frac{U_0 X}{L},0),
\label{vshr}
\end{eqnarray}
where $U_0$ is the speed at the two walls. Because of rotation, a
Coriolis acceleration acts on the fluid and is described by a
frequency
\begin{eqnarray}
\vec{\omega}=(0,0,\Omega_0),\hskip0.5cm\Omega_0=\frac{U_0}{q L}.
\label{om0}
\end{eqnarray}
Here the parameter $q$ is positive (corresponding to angular
velocity decreasing with increasing radius in a disk) and describes
the radial dependence of $\Omega(r)$ in the accretion disk,
\begin{eqnarray}
\Omega(r)=\Omega_0\left(\frac{r_0}{r}\right)^{q}.
\label{omrd}
\end{eqnarray}
Thus, $q=3/2$ corresponds to a Keplerian disk and $q=2$ corresponds
to a disk with a constant specific angular momentum.  For
completeness, we note that $q=1$ corresponds to a system with a flat
rotation curve and $q=0$ to solid body rotation.

The classical plane Couette flow that is widely discussed in the fluid
literature has a finite shear but no Coriolis force.  In our model,
this corresponds to a finite $U_0$ but zero $\Omega_0$, i.e., it
represents the limit $q\rightarrow \infty$.  The accretion disk
problem, which is of primary interest in astrophysics, corresponds to
finite $q$ in the range $3/2$ to $2$. In comparing the present work to
the fluid literature, the reader is warned that our radial coordinate
$X$ maps to $Y$ in the fluid work, while our $Y$ is their $X$.  The
notation we use is standard in the astrophysics literature.
Below we describe the self-contained set of generalized equations from beginning, 
for the convenience of the general reader.

\subsection{Perturbations}

The dynamics of a viscous incompressible fluid are described by the
Navier-Stokes equation (e.g., Landau \& Lifshitz 1989),
\begin{equation}
\frac{\partial {\vec V}}{\partial {t^\prime}}+{\vec V}.\nabla^\prime
{\vec V}+{\vec \omega}\times{\vec \omega}\times{\vec D}+2{\vec
\omega}\times{\vec V}+
\nabla^\prime\left(\frac{P}{\rho}\right)={\nu}{\nabla^\prime}^2{\vec
V},
\label{navst}
\end{equation}
supplemented with the condition of incompressibility,
\begin{equation}
\nabla^\prime.{\vec V}=0,
\label{eoc}
\end{equation}
where $t^\prime$ is time, ${\vec V}$ is the velocity, ${\vec \omega}$
is the Coriolis vector defined in equation (\ref{om0}), $\nu$ is the
kinematic coefficient of viscosity, $\vec{D}\equiv(X,Y,Z)$,
$\nabla^\prime\equiv\left({\partial/\partial X}, {\partial/\partial
Y},{\partial/\partial Z}\right)$, and $P$ is the pressure. Due to the
incompressibility assumption, the density $\rho$ is a constant.

It is convenient to analyse the perturbations in terms of
dimensionless variables, $x,y,z,t$, defined by
\begin{equation}
X=x L,\hskip0.1cm Y=y L,\hskip0.1cm Z=z L,\hskip0.1cm {\vec V}={\vec
U} U_0, \hskip0.1cm t^\prime=t L/U_0,
\label{dim}
\end{equation}
where ${\vec U}$ is a dimensionless velocity
\begin{equation}
\vec{U}=(0,U_y,0),\hskip0.5cm U_y=U(x)=-x.
\label{vdimls}
\end{equation}
Then, by substituting (\ref{dim}) into (\ref{navst}), we obtain
\begin{equation}
\frac{\partial {\vec U}}{\partial t}+{\vec U}.\nabla {\vec U}
+\frac{\hat{k}\times\hat{k}\times{\vec d}}{q^2}
+\frac{2\hat{k}\times{\vec U}}{q} +\nabla
\bar{p}=\frac{1}{R}{\nabla}^2{\vec U},
\label{navst2}
\end{equation}
where $\bar{p} U_0^2= P/\rho$, ${\vec d}\equiv(x,y,z)$,
$\nabla\equiv\left({\partial/\partial x}, {\partial/\partial
y},{\partial/\partial z}\right)$, and the Reynolds number $R$ is defined
by 
\begin{equation}
R={U_0 L \over \nu}.
\label{Reynolds}
\end{equation}

We consider small perturbations in the velocity components of the
form: $U_x\rightarrow u(x,y,z,t)$, $U_y\rightarrow U(x)+v(x,y,z,t)$,
$U_z\rightarrow w(x,y,z,t)$, and perturbations in the pressure
$\bar{p}\rightarrow \bar{p}+p(x,y,z,t)$.  The linearized Navier-Stokes
and continuity equation for the incompressible fluid then give
\begin{eqnarray}
\left(\frac{\partial}{\partial t}+U\frac{\partial}{\partial y}\right)u
-\frac{2v}{q}+\frac{\partial p}{\partial x}=\frac{1}{R}\nabla^2 u ,
\label{u}
\end{eqnarray}
\begin{eqnarray}
\left(\frac{\partial}{\partial t}+U\frac{\partial}{\partial y}\right)v
+u\frac{\partial U}{\partial x}+\frac{2u}{q}+\frac{\partial
p}{\partial y} =\frac{1}{R}\nabla^2 v ,
\label{v1}
\end{eqnarray}
\begin{eqnarray}
\left(\frac{\partial}{\partial t}+U\frac{\partial}{\partial y}\right)w
+\frac{\partial p}{\partial z}=\frac{1}{R}\nabla^2 w ,
\label{om}
\end{eqnarray}
\begin{equation}
\frac{\partial u}{\partial x}+\frac{\partial v}{\partial y}+
\frac{\partial w}{\partial z}=0.
\label{ec}
\end{equation}

Let us rewrite the equations in terms of the $x$ component of the
vorticity,
\begin{equation}
\zeta=\frac{\partial w}{\partial y}-\frac{\partial v}{\partial z}.
\label{vor}
\end{equation}
Combining equations (\ref{u})--(\ref{ec}) and simplifying,
we obtain
\begin{equation}
\nabla^2p=-2\frac{\partial U}{\partial x}\frac{\partial u}{\partial y}
+\frac{2}{q}\left(\frac{\partial v}{\partial x}-\frac{\partial
u}{\partial y}\right).
\label{p2}
\end{equation}
Eliminating $p$ and $v$ from (\ref{u}) by use of
(\ref{ec})--(\ref{p2}) we find
\begin{equation}
\left(\frac{\partial}{\partial t}+U\frac{\partial}{\partial y}\right)\nabla^2 u
-\frac{\partial^2U}{\partial x^2} \frac{\partial u}{\partial y}
+\frac{2}{q}\frac{\partial \zeta}{\partial z}
=\frac{1}{R}\nabla^4 u.
\label{v}
\end{equation}
Finally, combining (\ref{v1}) and (\ref{om}) by use of (\ref{vor}) we obtain
\begin{equation}
\left(\frac{\partial}{\partial t}+U\frac{\partial}{\partial y}\right)\zeta
-\frac{\partial U}{\partial x} \frac{\partial u}{\partial z}
-\frac{2}{q}\frac{\partial u}{\partial z}=\frac{1}{R}\nabla^2 \zeta,
\label{zeta}
\end{equation}
where we recall from equation (\ref{vdimls}) that $U=-x$.

Equations (\ref{v}) and (\ref{zeta}) are the standard Orr-Sommerfeld
and Squire equations, respectively, except that they now have
additional terms proportional to $2/q$ because of the inclusion of
Coriolis acceleration.  We are interested in solving these linear
equations with no-slip boundary conditions, i.e., $u=v=w=0$ at the two
walls. Equivalently 
\begin{equation}
u={\partial u \over \partial x}=\zeta=0,\hskip0.2cm{\rm at}\hskip0.1cm
{x=\pm 1}.
\label{bc}
\end{equation}

Because of translation-invariance of the unperturbed flow in $y$
and $z$, we can decompose the perturbations in terms of Fourier modes
in these directions.  Also, for convenience, we study the
perturbations in terms of $(u, \zeta)$ rather than
$(u,v,w)$. Therefore, we write the perturbations as
\begin{eqnarray}
\nonumber
u(x,y,z,t)=\hat{u}(x,t)\exp[i\vec{k}.\vec{r}_p],\\
\zeta(x,y,z,t)=\hat{\zeta}(x,t)\exp[i\vec{k}.\vec{r}_p],
\label{sol}
\end{eqnarray}
where $\vec{r}_p$ [$\equiv(y,z)$] is any radius vector in the $y-z$
plane and $\vec{k}\equiv(k_y,k_z)$.  By substituting (\ref{sol}) into
(\ref{v}) and (\ref{zeta}), we obtain
\begin{eqnarray}
\nonumber
\frac{\partial \hat{u}}{\partial t}&=&-i[{\cal L}_{os} \hat{u}
+{\tilde{\cal L}}_{cor}\hat{\zeta}] ,\\
\frac{\partial \hat{\zeta}}{\partial t}&=&-i[({\cal L}_{c}+{\cal L}_{cor})
\hat{u}+{\cal L}_{sq} \hat{\zeta}] ,
\label{sol2}
\end{eqnarray}
where
\begin{eqnarray}
\nonumber
{\cal L}_{os}&=&-(D^2-k^2)^{-1}[(D^2-k^2)^2/(iR)-k_y U(D^2-k^2)+k_y D^2U],\\
\nonumber
{\cal L}_{c}&=&-k_z D U,\\
\nonumber
{\cal L}_{sq}&=&k_y U-(D^2-k^2)/(iR),\\
\nonumber
{\cal L}_{cor}&=&-\frac{2k_z}{q},\\
\nonumber
\tilde{{\cal L}}_{cor}&=&-{\cal L}_{cor}(D^2-k^2)^{-1},\\
D&=&\partial/\partial x.
\label{op}
\end{eqnarray}
If we further define
\begin{equation} 
Q=\left(\begin{array}{cr} \hat{u}\\ \hat{\zeta}\end{array}\right),
\qquad {\cal L}=\left(\begin{array}{cr} {\cal L}_{os} & {\tilde{\cal
L}}_{cor}\\ {\cal L}_c+{\cal L}_{cor} & {\cal L}_{sq}
\end{array}\right),
\label{arrays}
\end{equation}
equation (\ref{sol2}) reduces to the form
\begin{eqnarray}
\frac{\partial Q}{\partial t}=-i{\cal L} Q,
\label{sol3}
\end{eqnarray}
which we need to solve to obtain the eigenvectors and corresponding
eigenvalues. Since the set of eigenmodes for this bounded flow problem
is discrete and complete, we can write the solution to (\ref{sol3}) in
terms of an eigenfunction expansion,
\begin{eqnarray}
\nonumber
Q(x,t)&=&\sum_{j=1}^{\infty} \left[A_j \exp(-i\lambda_j t) {\tilde Q}^1_j(x)+B_j \exp(-i\mu_j t)
{\tilde Q}^2_j(x)\right],\\
\nonumber
\tilde{Q}^1_j(x)&=&\left(\begin{array}{cr}\tilde{u}^1_j(x)\\
\tilde{\zeta}_j^1(x)\end{array}\right),\\
{\tilde Q}^2_j(x)&=&\left(\begin{array}{cr} \tilde{u}^2_j(x)\\
{\tilde \zeta}^2_j(x)\end{array}\right),
\label{sol4}
\end{eqnarray}
where $(\lambda_j,{\tilde Q}^1_j(x))$ is the Orr-Sommerfeld
eigensystem\footnote{A complete set of eigenvalues and eigenvectors
is called the eigensystem.} and $(\mu_j,{\tilde Q}^2_j(x))$ is the Squire eigensystem.
Formally merging the two systems, we can rewrite (\ref{sol4}) as
\begin{eqnarray}
\nonumber
Q(x,t)&=&\sum_{j=1}^{\infty} C_j \exp(-i\sigma_j t) {\tilde Q}_j(x),\\
{\tilde Q}_j(x)&=&\left(\begin{array}{cr}\tilde{u}_j(x)\\
\tilde{\zeta}_j(x)\end{array}\right),
\label{sol5}
\end{eqnarray}
where half the indices $j$ correspond to the Orr-Sommerfeld modes and the 
other half of $j$ correspond to the Squire modes, and 
$\sigma_j=\sigma_{Rj}+i\sigma_{Ij}$. Therefore, for the $j$th
mode, (\ref{sol3}) reduces to
\begin{eqnarray}
{\cal L} {\tilde Q}_j=\sigma_j {\tilde Q}_j.
\label{sol6}
\end{eqnarray}
To calculate the set of eigenvalues and eigenvectors, we convert the
differential operator ${\cal L}$ into an $N\times N$ matrix in a
finite-difference representation and we then compute the eigenvalues
and eigenvectors of the matrix. The required order $N$ of the matrix
for adequate accuracy depends on the physical parameters of the
problem (mainly $R$ and also $k_y,k_z$). For the calculations
presented here, we used $N$ in the range $200-300$ (i.e., each of the
blocks ${\cal L}_{os}$, ${\cal L}_{sq}$, ${\cal L}_c$ and ${\cal
L}_{cor}$ had a size in the range $100$ to $150$).  

\subsection{Energy Growth}

\begin{figure}
\epsscale{1.0}
\plotone{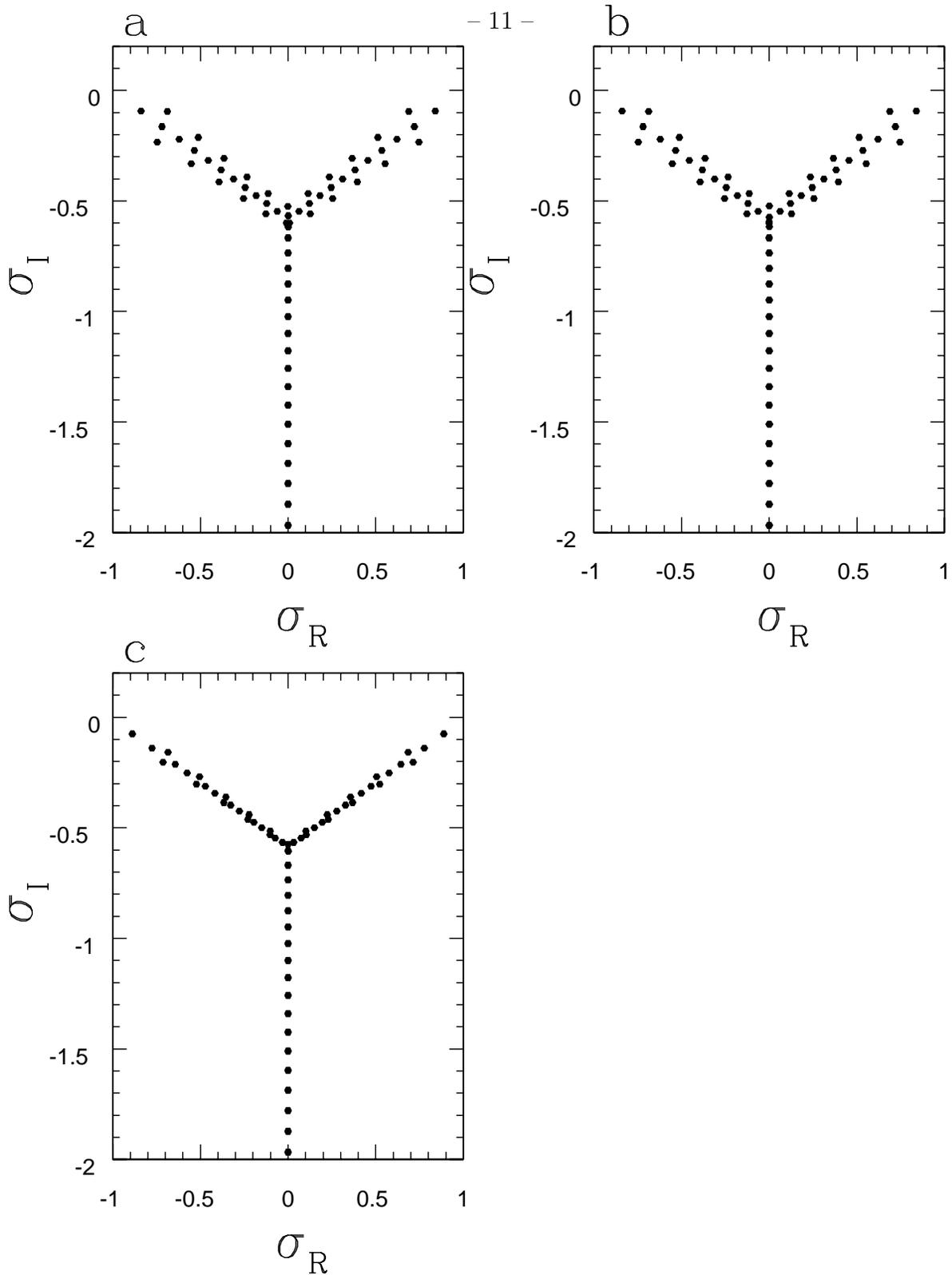}
\vskip3.5cm
\caption{
\label{fig1} 
Eigenspectra of (a) plane Couette flow, (b) constant angular momentum
disk ($q=2$), and (c) Keplerian disk ($q=3/2$), for $k_y=k_z=1$,
$R=2000$.  }
\end{figure}

The eigenvalues $\sigma_j$ and the corresponding eigenvectors ${\tilde
Q}_j$ for plane Couette flow have been studied by a number authors
(e.g. Orszag 1971; Romanov 1973; Farrell 1988; Reddy \& Henningson
1993) who have shown that there are no exponential growing eigenmodes
in the system.  That is, for no choice of the parameters is there an
eigenvalue with {\it positive} $\sigma_{I}$.  Figure \ref{fig1}a shows
such a typical eigenspectrum: the case shown has $R=2000$ and
$k_y=k_z=1$. For comparison Fig. \ref{fig1}b shows the eigenspectrum of a disk
with constant angular momentum $q=2$, and Fig. \ref{fig1}c a
Keplerian disk $q=3/2$. To the best of our knowledge,
this one to one comparison of eigenspectra between standard plane Couette flow,
a constant angular momentum flow and a Keplerian flow has not been reported 
earlier.  It is clear that none of
these flows has any growing eigenmode and so all three systems are
linearly stable. It should also be noticed that the eigenspectrum for
plane Couette flow is very similar to that of a constant angular momentum
flow. We explore this similarity further in \S\S3,4.


One of the most important features of plane Couette flow (and its
governing linear operator) is that the eigenmodes of the system are
close to linearly dependent in nature i.e. they are {\it non-normal} in nature.  Because
of this, even in the absence of any exponentially growing mode in the
system, it is possible to have a large transient growth in the energy
of certain perturbations (Butler \& Farrell 1992; Trefethen et
al. 1993). This growth occurs in the absence of non-linear effects and
is believed to play an important role in the transition from laminar
to turbulent flow.

Following previous authors (e.g. Trefethen et al. 1993; Schmid \&
Henningson 1994) we define the perturbed energy density as
\begin{equation}
E=\frac{1}{2V}\int_{-1}^{1}\int_0^a\int_0^b (u^2+v^2+w^2)dzdydx,
\label{energy}
\end{equation}
where $a=2\pi/k_y$, $b=2\pi/k_z$, and $V=2ab$ is the integration
volume.  We then seek to maximize the growth in this quantity.
Recalling the formal solution of (\ref{sol3}) in matrix form,
\begin{eqnarray}
Q(x,t)=\exp[-i{\cal L}t]Q(x,0),
\label{solmat}
\end{eqnarray}
the maximum growth in the perturbed energy can be expressed as
\begin{eqnarray}
G(k_y,k_z,R,t)={\rm
maximum}\left(\frac{||Q(^{.},t)||^2_2}{||Q(^{.},0)||^2_2}\right)=||\exp[-i{\cal
L}t]||^2_2,
\label{grow1}
\end{eqnarray}
where $||...||_2$ signifies the {\it norm} of the respective quantity and
the subscript $2$ specifies the 2-norm or Euclidian norm. The 2-norm of the
matrix can be evaluated by means of a singular value
decomposition. Then, for a given $t$, the square of the highest singular value is the maximum
energy growth, $G_{max}(t)$, for that time.
Physically, by ``maximum'' we mean that we consider all possible initial
perturbations $Q(x,0)$ and choose that function that maximizes the
growth of energy at time $t$. The corresponding energy growth factor
is $G(k_y,k_z,R,t)$.  When $t=0$, by definition $G(k_y,k_z,R,t)=1$,
implying no growth.  For given $t$, we maximize $G(k_y,k_z,R,t)$ by
writing $Q(x,0)$ as a linear combination of the eigenmodes of the
system as in equation (\ref{sol5}) and optimizing the coefficients
$C_j$. The details are given in the Appendix.


We should mention that to evaluate the growth one does not need to
include all the eigenmodes in the computation.  It has been shown by
Reddy \& Henningson (1993) that only a limited number, $K/2$, of the
Orr-Sommerfeld and Squire modes, viz., those with the largest (i.e.,
least negative) $\sigma_I$ values, are responsible for the growth.
The remaining modes decay too rapidly to provide much growth.
Therefore (\ref{sol5}) can be rewritten as
\begin{eqnarray}
Q_K(x,t)=\sum_{j=1}^{K} C_j \exp(-i\sigma_j t) {\tilde Q}_j(x)
=\tilde{Q}\exp[-i\Sigma_K t]C
\label{sol8}
\end{eqnarray}
and the corresponding growth is $G_K(k_y,k_z,R,t)$.
Here $\tilde{Q}$ and $C$ are $N\times K$ and $K\times 1$ matrices
respectively and $\Sigma_K$ is a $K\times K$ diagonal matrix
consisting of the top $K$ eigenvalues (top $K/2$ of Orr-Sommerfeld and
$K/2$ of Squire eigenvalues).  For the calculations presented in this
paper, we generally used $K\le 60$.

The growth $G(k_y,k_z,R,t)$ defined above is a function of four
parameters.  In various places in the paper we consider different
kinds of maxima of this function.  For instance, for fixed
$k_y,k_z,R$, we could maximize $G$ with respect to time $t$, 
and thereby determine the maximum growth
$G_{max}(k_y,k_z,R)$. This is the quantity
that is plotted as contours in Figures 2 and 4.  Or, we may wish to
hold one of the components of the wavevector fixed, e.g., $k_y=0$
(see Table 1, \S4.1, also Figure 6 for other values of $k_y$) or $k_z=0$
(\S4.2), and optimize the growth with respect to the other component
of the wavevector and the time; this gives maximum growth factors such
as $G_{max}(k_y=0,R)$ and $G_{max}(k_z=0,R)$.  Finally, for a given
$R$, we could optimize over all the other parameters to
calculate $G_{max}(R)$.  This is the quantity of most interest, and is
shown for instance in Tables 1, 2, and Figures 3, 5.

\section{Numerical Results}

\subsection{Plane Couette Flow and Constant Specific Angular Momentum Flow}

\begin{figure}
\epsscale{0.8}
\plotone{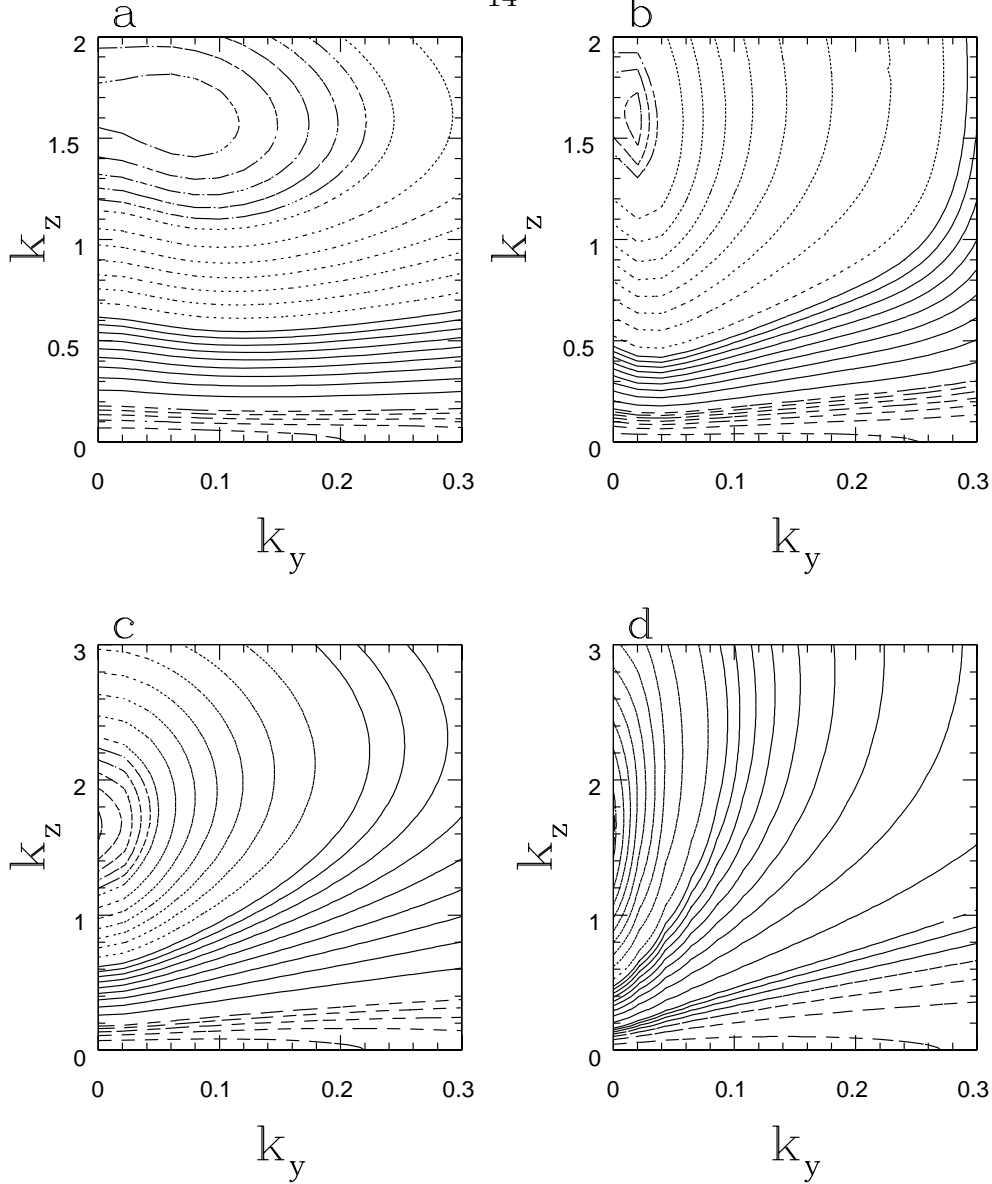}
\vskip1.05cm
\caption{
\label{fig2} 
Contours of $G_{max}(k_y,k_z,R)$ in the $k_y-k_z$ plane.  (a) Plane
Couette flow for $R=500$: dashed contours correspond to
$G_{max}=2,4,...,10$, solid contours to $G_{max}=20,30,...,100$,
dotted contours to $G_{max}=120,140,...,240$, and dot-dashed contours
to $G_{max}=250,260,...,290$.  (b) Plane Couette flow for $R=2000$:
dashed contours correspond to $G_{max}=10,30,...,130$, solid contours
to $G_{max}=200,300,...,1000$, dotted contours to
$G_{max}=1200,1600,...,4000$, and dot-dashed contours to
$G_{max}=4500,4600,4700$.  (c) Constant angular momentum disk ($q=2$)
for $R=500$: dashed contours correspond to $G_{max}=2,4,...,10$, solid
contours to $G_{max}=20,30,...,100$, dotted contours to
$G_{max}=120,140,...,240$, and dot-dashed contours to
$G_{max}=250,260,...,290$.  (d) Constant angular momentum disk for
$R=2000$: dashed contours correspond to $G_{max}=10,30,...,130$, solid
contours to $G_{max}=200,300,...,1000$, dotted contours to
$G_{max}=1200,1600,...,4000$, and dot-dashed contours to
$G_{max}=4500,4600$.  }
\end{figure}

In the previous section, we showed that the eigenspectra of plane
Couette flow and a constant angular momentum disk ($q=2$) are very
similar. Here we show that the maximum growths are also
similar. Figures \ref{fig2}a,b and \ref{fig2}c,d show contours of
constant $G_{max}$ in the $\{k_y,k_z\}$ plane for plane Couette flow
and a $q=2$ disk respectively for two values of $R$ ($500,2000$).  The maximum
growth values for these two cases and for other values of $R$ are
found in Fig. 3.  We see that the values are very similar for the two
flows.  Whereas for $q=2$ the maximum growth occurs exactly on the
$k_z$ axis ($k_y=0$), for plane Couette flow it is slightly off the
axis, though by a progressively smaller amount with increasing $R$.
This deviation in location of the occurrence of maximum growth in the $k_y-k_z$ plane 
for a constant angular momentum disk compared to plane Couette flow was
completely unnoticed earlier, to the best of our knowledge. In cases of
a large $R$, the best growth values for $k_y=0$ compared to $k_y\sim 0$ (but $\neq 0$)
do not have any practical difference. But for a small $R$ the difference is 
important as in this case the maximum growth factor itself is small (see Fig. 3).
Therefore, in presence of finite molecular viscosity this result has a 
physical implication to fluid dynamics, though in the accretion disk
when $R$ is always expected to be very large this may not be an important issue.

\begin{figure}
\epsscale{.80}
\plotone{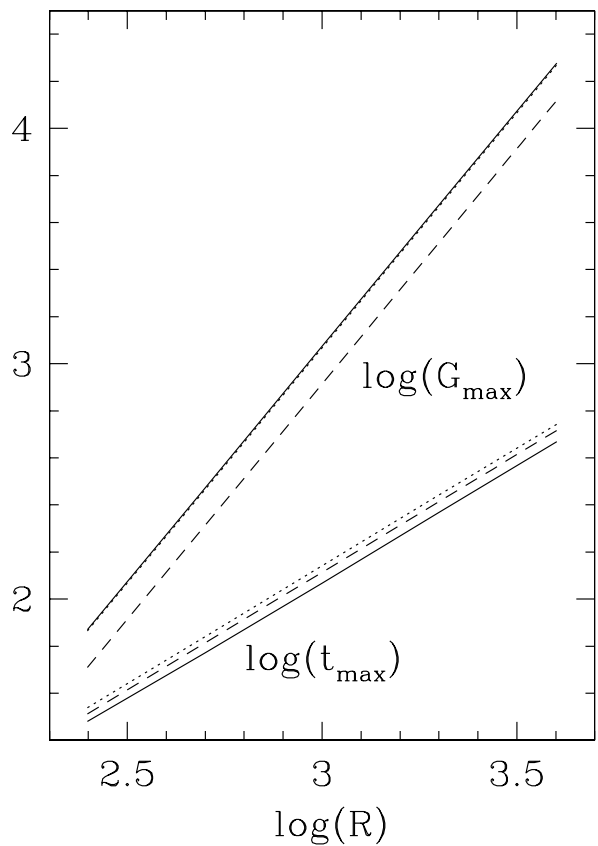}
\vskip1.5cm
\caption{
\label{fig3} 
$G_{max}(R)$ and $t_{max}(R)$ as functions of Reynolds number
$R$. Solid lines correspond to plane Couette flow and dotted lines to
a constant angular momentum disk ($q=2$).  The dashed lines show the
analytic result discussed in \S 4.1.1. }
\end{figure}

In Figure \ref{fig3}, we show the variation of the maximum
growth $G_{max}$ and the corresponding time $t_{max}$ at which the
maximum growth occurs as functions of $R$ for plane Couette flow and
$q=2$.  We see that $G_{max}$ varies as $R^2$ and $t_{max}$ as $R$ in
both cases, with very similar values, again indicating the similarity
of the two flows. However for plane Couette flow, growth maximizes for
$k_z\sim 1.6$, while for a $q=2$ disk it happens at $k_z=1.66$. Moreover,
in plane Couette Flow, the optimum $k_y$ scales as $1/R$, already noticed
by earlier authors (e.g. Butler \& Farrell 1992), whereas for a $q=2$
flow the optimum $k_y=0$.

In the case of a constant angular momentum disk the epicyclic
frequency of the disk becomes zero which makes the basic structure of
the system very similar to that of plane Couette flow (compare the
equation set (\ref{u})-(\ref{om}) as well as (\ref{v}) and
(\ref{zeta}) for plane Couette  flow and constant angular momentum
flow). This was already noticed by Balbus, Hawley \& Stone (1996) and Hawley, 
Balbus \& Winters (1999), who found from numerical simulations that these two flows
are equally susceptible to hydrodynamic turbulence.  We explore the
physics of this similarity further in \S4.

\clearpage

\subsection{Effect of Non-Zero Epicyclic Frequency}

The epicyclic frequency is given by
\begin{equation}
\kappa = \sqrt{2(2-q)} \Omega,
\label{epicyclic}
\end{equation}
which is zero for $q=2$ and is non-zero (and real) for any $q<2$.
Figures {\ref{fig2}(c),(d) show that, when $q=2$, the maximum growth occurs for
axisymmetric perturbations with vertical structure (hereafter we refer as
``vertical perturbation"), i.e. $k_y=0$, $k_z
\ne 0$.  We begin this section by exploring what happens to such
vertical perturbations when $q<2$. Note that most of the fluid literature is devoted to
plane Couette flow with no Coriolis force. In the astrophysics literature,
Yecko (2004) mainly concentrated on the Keplerian
disk ($q=1.5$) and plane Couette flow, whereas we consider the full range, $1.5\leq q\leq 2$.

Table 1 shows the maximum growth for vertical perturbations for four
values of $q$: 1.99, 1.9, 1.7, 1.5.  We see that, as
$q$ decreases, the growth falls dramatically, and so does the time at
which the maximum occurs.  For a Keplerian flow, the growth factor is under
4.  Moreover, numerical experiments show that the growth is
insensitive to the value of $R$.  In other words, the growth is not
limited by viscosity but rather by the dynamics itself.  We explore the
reasons for this in \S4.1. Similar results are discussed
in AMN05, using a Lagrangian picture.

\vskip0.2cm {\centerline{\large Table 1}} {\centerline{Energy
Growth Factors of Disks for $R=2000$ and $k_y=0$}}
\begin{center}
{
\vbox{
\begin{tabular}{ccccllllllllllllll}
\hline
\hline
$q$ & $k_z$ &  $G_{max}(k_y=0,R=2000)$ & $t_{max}(k_y=0,R=2000)$    \\
\hline
\hline
$1.5$  & $2.5$ & $3.84$   & $2.8$  \\
\hline
$1.7$ & $2.4$ & $6.31$   & $4.1$  \\
\hline
$1.9$  & $2.1$ & $18.16$   & $8.3$  \\
\hline
$1.99$  & $1.9$ & $153.4$   & $27.8$  \\
\hline
 $2$  & $1.66$ & $4661$   & $277$  \\
\hline
\hline
\end{tabular}
}}
\end{center}

We next remove the restriction to vertical perturbations and consider
general $k_y$, $k_z$.  Figure \ref{fig4}a shows contours of constant
growth for $q=1.99$ for $R=2000$.  Even though the value of $q$ is
only very slightly different form that used in Figure \ref{fig2}d
($q=2$), nevertheless we see a dramatic change.  The main qualitative
difference between the two cases is that the epicyclic frequency is
zero for $q=2$ but is (slightly) non-zero for $q=1.99$.  Still, this small changes
causes a major modification of the results, showing what a dominant
effect the epicyclic frequency has on the fluid dynamics.  The other
panels in Fig. \ref{fig4} show results for other values of $q$.  It is
interesting to see how the location of the maximum in the $k_y-k_z$
plane changes as the system approaches the Keplerian regime, and also
how the magnitude of the growth reduces. This change in location of 
maximum growth as a function of $q$ in the $k_y-k_z$ plane has not been
noted earlier, to the best of our knowledge. For $q=2$ the maximum energy
growth is a factor of $4600$ and occurs on the $k_y=0$ line
(Fig. \ref{fig2}d) while for $q=1.5$ the maximum growth is only $22$
and occurs on the $k_z=0$ line (Fig. \ref{fig4}d).  Therefore for a
constant angular momentum disk, we need to include vertical structure
in the perturbations to maximize the energy growth, whereas for a
Keplerian disk a 2-dimensional analysis is sufficient.  Table 2 lists
the maximum energy growth factors for $q=1.5-2$ when $R=2000$.

Figures \ref{fig5}a and \ref{fig5}b show respectively the variation of
the maximum growth and the time at which the maximum growth occurs as
a function of Reynolds number in a Keplerian disk.  It is seen that
for large $R$, $G_{max}$ scales as $R^{2/3}$ and $t_{max}$ as
$R^{1/3}$.  This suggests that, even though the growth is modest for
the values of $R$ we have considered, if we go to sufficiently large
values of $R$, very large energy growth might still be possible.  This
is of interest because the Reynolds number of a cold accretion disk is
very high (many orders of magnitude higher than the values considered
in this work), so turbulence could be generated in such systems.  Due
to numerical constraints our current results are limited to $R\le
10^4$; however, this range captures most of the basic features of the
growth. Yecko (2004) used a superior spectral code and was able to
go to much larger values of $R$.

Figure \ref{fig6} shows how $G_{max}$ and $t_{max}$ scale with $k_y$
at a given $R$.  The maximum growth is achieved at $k_y\sim1.2$.  At
smaller $k_y$, $G_{max}$ scales as $k_y^{2/3}$, while at larger
$k_y$, $G_{max}$ decreases as $\sim 1/k_y$ or $k_y^{-4/3}$.  Also
$t_{max}$ scales as $k_y^{-2/3}$ at large $t$. 
Yecko (2004) and Umurhan \& Regev (2004) identified the 
scaling of $G_{max}$ with $R$ for a Keplerian disk, but the other scaling
relations have not been discussed before.
We also derive all the scaling relations analytically, for the first time, 
in \S4. A detailed
understanding of these scalings is given in \S 4.2.
Identical scaling relations are also derived by AMN05.

\clearpage
\vskip0.2cm
{\centerline{\large Table 2}}
{\centerline{Maximum Energy Growth for Disks with Various Values of $q$ and $R=2000$}}
\begin{center}
{
\vbox{
\begin{tabular}{ccccclllllllllllllll}
\hline
\hline
$q$ & $k_y$ & $k_z$ &  $G_{max}(R=2000)$ & $t_{max}(R=2000)$    \\
\hline
\hline
$1.5$  & $1.2$ & $0$ & $21.67$   & $11$  \\
\hline
$1.7$ & $1.05$ &$0.6$ & $23$   & $12.4$  \\
\hline
 $1.9$  & $0.34$ &$0.96$ & $32.33$   & $19.8$  \\
\hline
$1.99$  & $0.14$ &$1.64$ & $174.43$   & $34.3$  \\
\hline
 $2$  & $0$ &$1.66$ & $4661$   & $277$  \\
\hline
\hline
\end{tabular}
}}
\end{center}

\begin{figure}
\epsscale{0.8}
\plotone{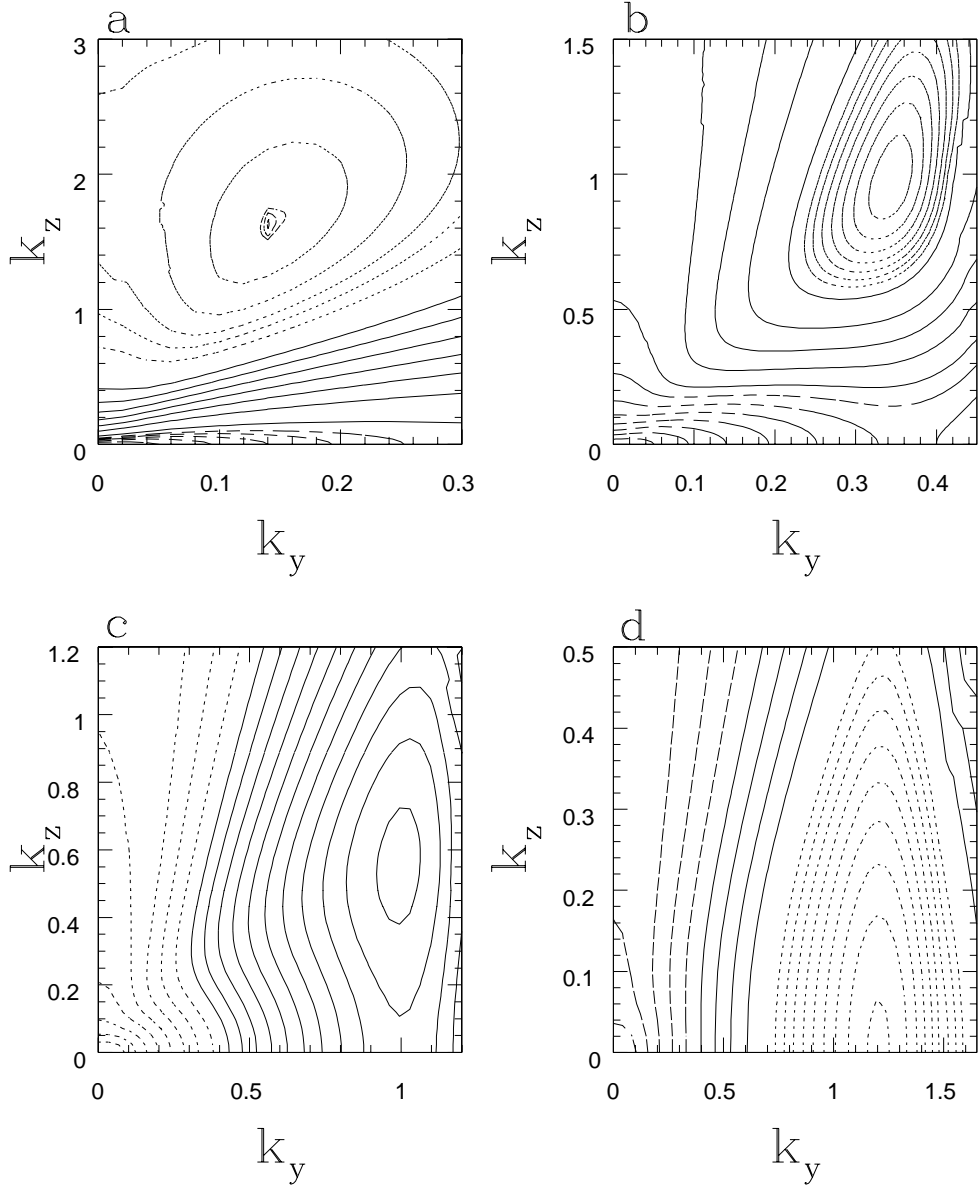}
\vskip1.7cm
\caption{
\label{fig4} 
Contours of $G_{max}(k_y,k_z,R)$ for disks with different values of
$q$ and $R=2000$. (a) $q=1.99$: dashed contours correspond to
$G_{max}=2,4,...,10$, solid contours to $G_{max}=15,30,...,105$,
dotted contours to $G_{max}=130,140,...,170$, and dot-dashed contours
to $G_{max}=174,174.2,174.4$.  (b) $q=1.9$: dashed contours correspond
to $G_{max}=2,4,...,14$, solid contours to $G_{max}=16,19,...,28$, and
dotted contours to $G_{max}=29,29.5,...,32$.  (c) $q=1.7$: dotted
contours correspond to $G_{max}=2,3,...,9$, and solid contours to
$G_{max}=11,12.2,...,23$.  (d) $q=1.5$: dashed contours correspond to
$G_{max}=2,3,...,6$, solid contours to $G_{max}=7,9,...,13$, and
dotted contours to $G_{max}=17.5,18,...,21.5$.  }
\end{figure}

\begin{figure}
\epsscale{1.0}
\plotone{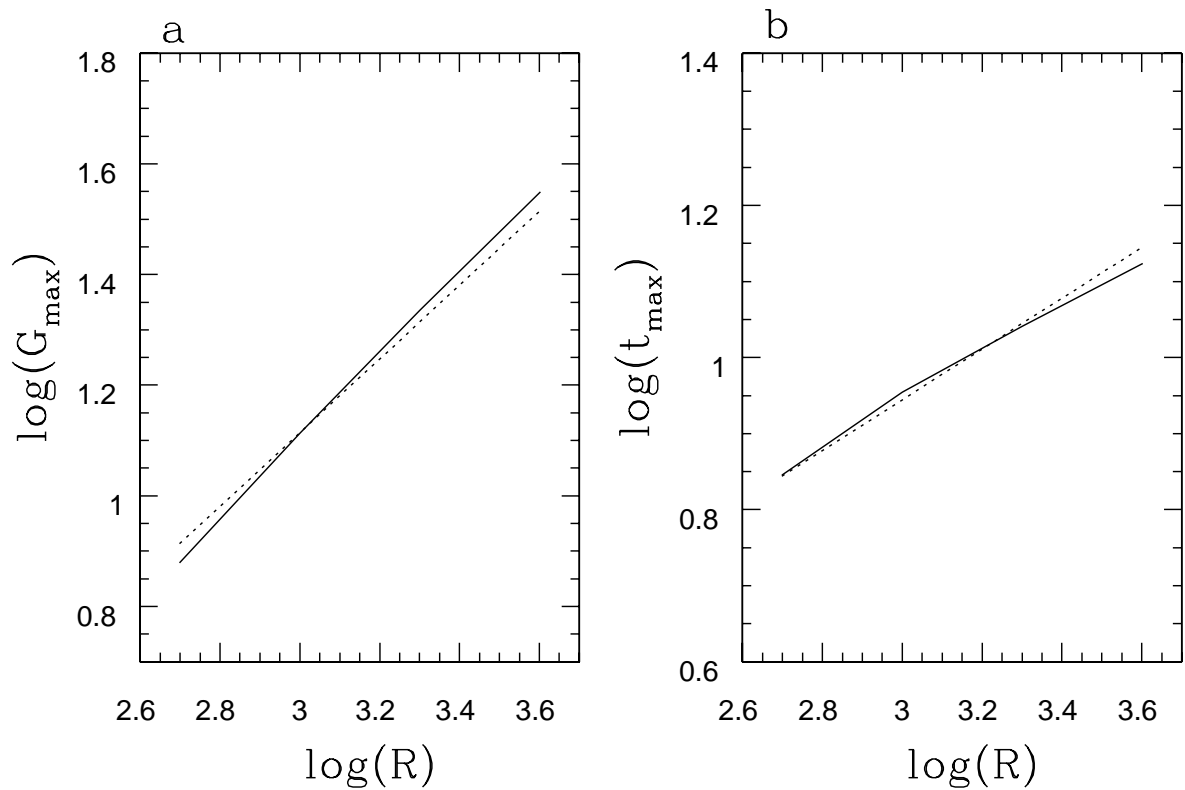}
\vskip-6.5cm
\caption{
\label{fig5} 
Solid lines show (a) $G_{max}(R)$, and (b) $t_{max}(R)$, as functions
of $R$ for $q=1.5$.  Dotted lines correspond to the analytical result
for $k_{x,min}=1.7$ discussed in \S4.2.
}
\end{figure}

\begin{figure}
\epsscale{.50}
\plotone{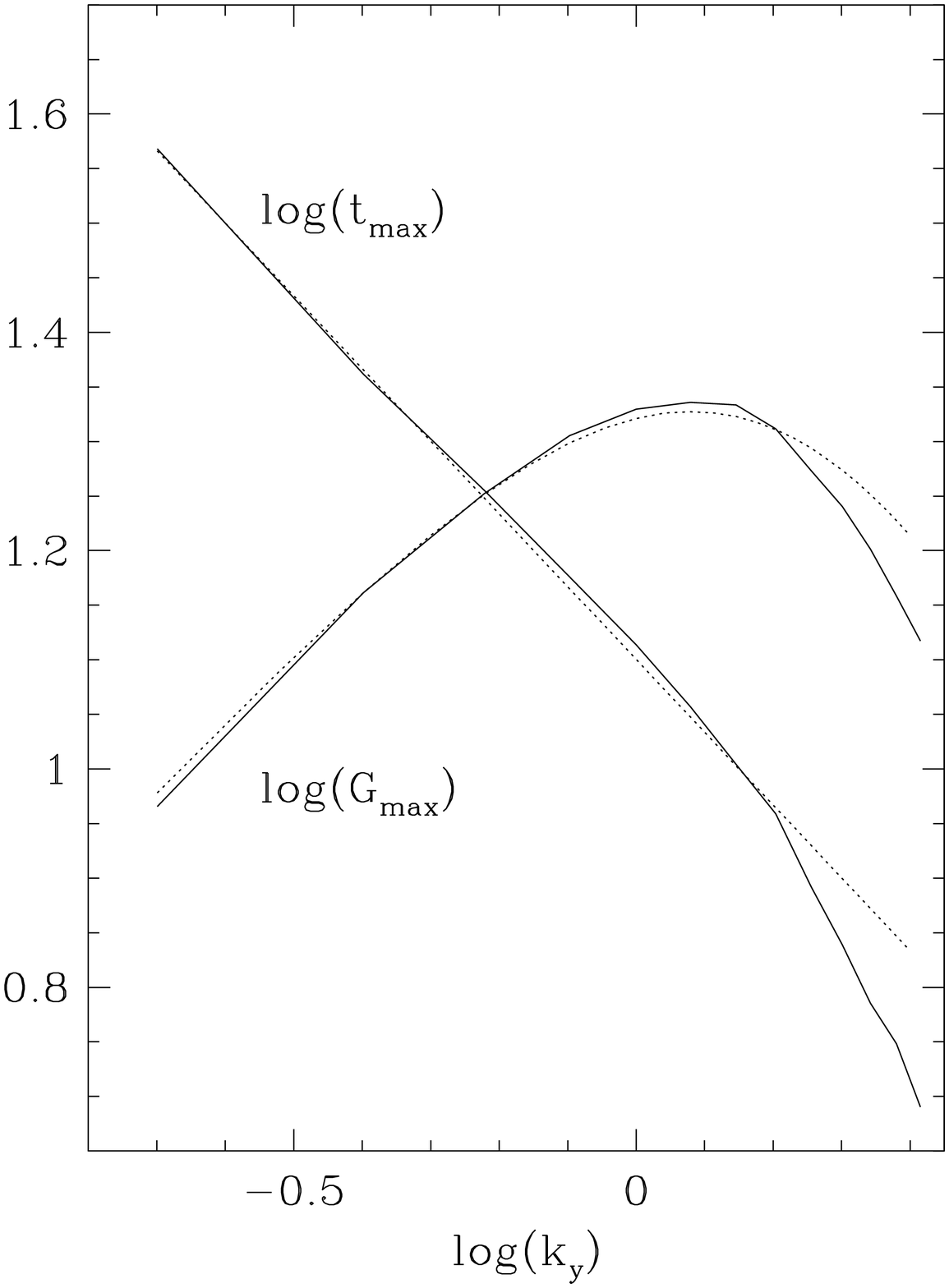}
\vskip1.5cm
\caption{
\label{fig6} 
$G_{max}(k_y,R)$ and $t_{max}(k_y,R)$ as functions of $k_y$ for
$q=1.5$ and $R=2000$.  The dotted lines correspond to the analytical
result discussed in \S 4.2.}
\end{figure}

\subsection{Nature of the Growing Perturbations in a Keplerian Flow}

We have seen that for a Keplerian flow the maximum growth occurs for
$k_y\sim 1.2$, $k_z=0$. Figures \ref{fig7} and \ref{fig8} show the
development with time of the perturbed velocity component $u(x,y)$
corresponding to $R=500,4000$, respectively, optimized for the best
growing mode. In each case, we show
snapshots corresponding to four times: $t=0$, $t_{max}/2$, $t_{max}$,
$3t_{max}/2$.  The perturbations are seen to resemble plane waves that
are frozen in the shearing flow.  The initial perturbation at $t=0$ is
a leading wave with negative $k_x$ and with $|k_x| \gg k_y$.  With
time, the wavefronts are straightened out by the shear, until at
$t=t_{max}$, the wavefronts are almost radial and $k_x\sim0$.  At yet
later times, the wave becomes trailing and the energy also decreases.
The perturbations are very similar to the growing perturbation
described by Chagelishvili et al. (2003) and Umurhan \& Regev
(2004). However, those authors (and also AMN05) considered an infinite system whereas
our fluid is confined to a box of size $2L$ in the $x$ direction.
Figure \ref{fig9} shows the optimum growth of the energy $G(t)$ as a function of time
for the two perturbations whose time evolution for the 
best growing modes are shown in Figures 7 and 8. 

\begin{figure}
\epsscale{0.7}
\plotone{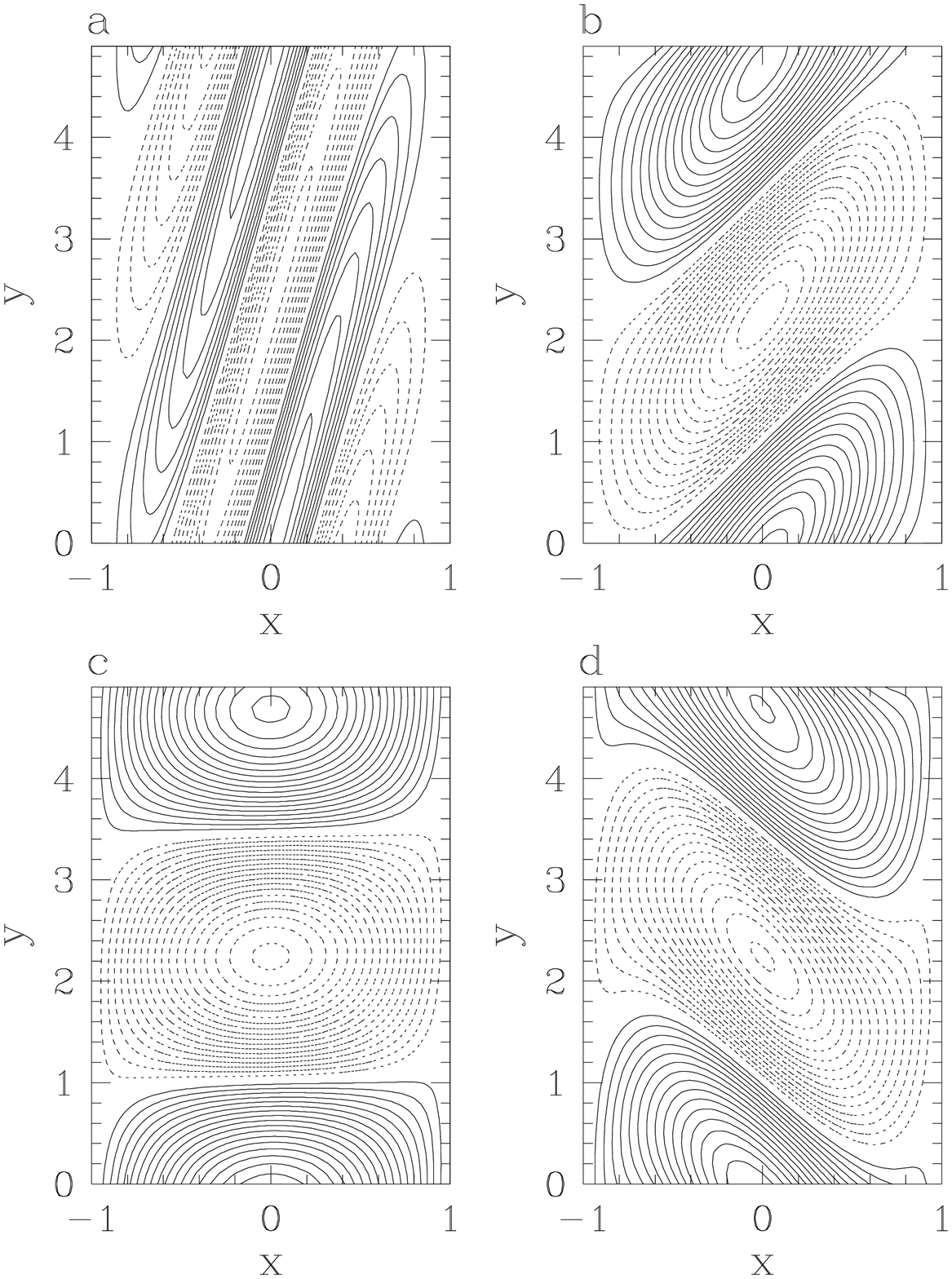}
\vskip1.9cm
\caption{
\label{fig7} 
Shows the development of the perturbed velocity $u(x,y)$ as a function
of time for the best growing energy optimal in a Keplerian flow
with $R=500$.  The perturbation has $k_y=1.29$, $k_z=0$, and the
maximum growth is achieved at $t_{max}(R=500)=6.6$.  The four panels
correspond to (a) $t=0$, (b) $t=t_{max}/2=3.3$, (c) $t=t_{max}=6.6$,
(d) $t=3t_{max}/2=9.9$.  Solid and dotted contours correspond to
positive and negative values of $u$ respectively.  }
\end{figure}

\begin{figure}
\epsscale{0.7}
\plotone{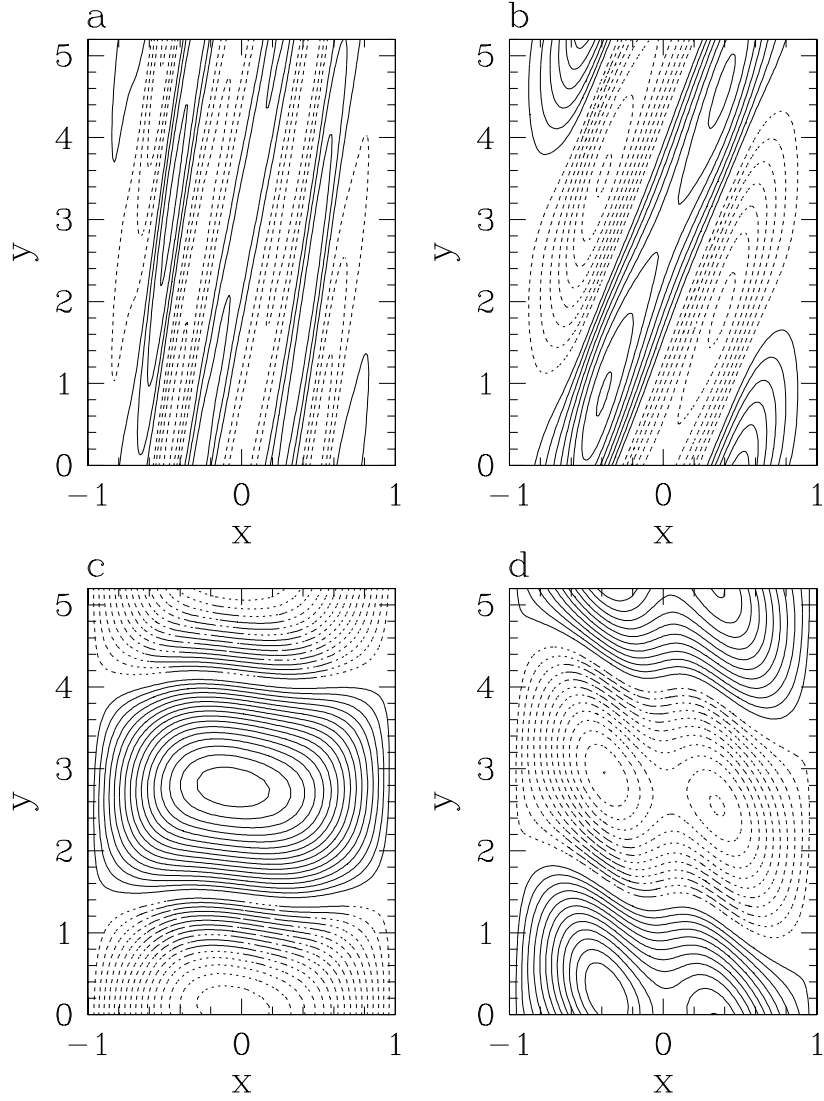}
\vskip1.9cm
\caption{
\label{fig8} 
Same as Fig. \ref{fig7} but for $R=4000$.  Here $k_y=1.2$, $k_z=0$,
$t_{max}(R=4000)=13.3$, and the four panels correspond to (a) $t=0$, (b)
$t=t_{max}/2=6.65$, (c) $t=t_{max}=13.3$, (d) $t=3t_{max}/2=19.95$.
Solid and dotted contours correspond to positive and negative values of
$u$ respectively.  }
\end{figure}

\begin{figure}
\epsscale{.5}
\plotone{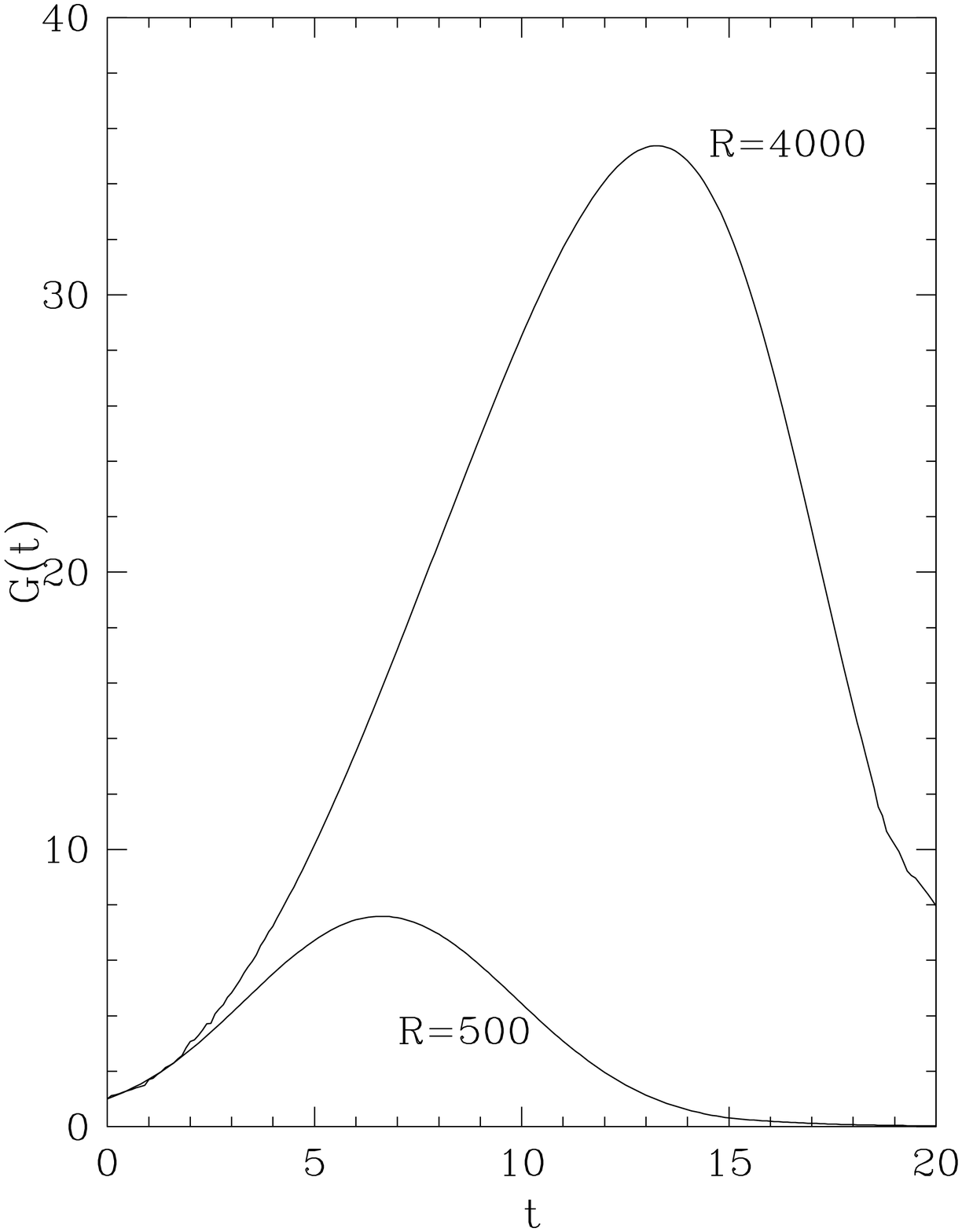}
\vskip1.5cm
\caption{
\label{fig9} 
Growth of the perturbed energy $G(t)$ as a function of time for the
cases shown in Figs. \ref{fig7} and \ref{fig8}.  }
\end{figure}

\section{Physical Interpretation of the Numerical Results}

In this section we attempt to understand via an analytical approach
the numerical results of the previous sections. We also like to derive 
the scaling relations described in the previous sections analytically.
We show that the analytical
solutions match the numerical results well. In the interest of
clarity, we work with the original dimensioned equations.  Thus $X$
(going from $-L$ to $+L$), $Y$, $Z$ are our coordinates, and we write
the corresponding components of the wavevector as $k_X$, $k_Y$, $k_Z$,
respectively.  We use $t$ for time (called $t'$ in \S 2.1).

Let us define the shear frequency $2A$ and the vorticity frequency
$2B$ as follows (Narayan, Goldreich \& Goodman 1987),
\begin{equation}
2A = -q\Omega, \qquad 2B = 2(A+\Omega) = (2-q)\Omega.
\label{freqs}
\end{equation}
Then the three components of the momentum equation and the
incompressibility condition give
\begin{equation}
{du\over dt} = 2\Omega v - {\partial \hat{p}\over \partial X}
+\nu {\nabla^\prime}^2 u,
\label{xmmtm}
\end{equation}
\begin{equation}
{dv\over dt} = -2B u - {\partial \hat{p}\over \partial Y}
+\nu {\nabla^\prime}^2 v,
\label{ymmtm}
\end{equation}
\begin{equation}
{dw\over dt} =  - {\partial \hat{p}\over \partial Z}
+\nu {\nabla^\prime}^2 w,
\label{zmmtm}
\end{equation}
\begin{equation}
{\partial u\over\partial X} + {\partial v\over\partial Y} + {\partial
 w\over\partial Z} = 0, \end{equation} where $\hat{p}=P/\rho$.  The
 Lagrangian time derivative $d/dt$ is given by
\begin{equation}
{d\over dt} = {\partial\over\partial t} -q\Omega X
{\partial\over\partial Y}.
\end{equation}

The numerical results described in \S3 showed that plane Couette flow
and $q=2$ flow both have maximum energy growth for perturbations with
vertical structure, whereas Keplerian $q=3/2$ flow has maximum growth
for two-dimensional perturbations with no vertical structure.  We
analyse these two cases separately.

\subsection{Vertical Perturbations}

To begin with, let us ignore the walls and assume plane wave solutions
of the form
\begin{equation}
u(X,Y,Z,t) = u(t) \exp(ik_XX+ik_YY+ik_ZZ),~{\rm etc.}.
\end{equation}
Furthermore, since perturbations with $k_Y=0$ (equivalent to $k_y=0$)
were seen to grow robustly for both plane Couette flow and $q=2$, let
us assume $k_Y=0$.  For such perturbations, $d/dt=\partial/\partial t$.
 
\subsubsection{Plane Couette Flow}

For plane Couette low, we set $2\Omega=0$ in equation (\ref{xmmtm})
and $2B=2A$ in equation (\ref{ymmtm}).  We can show that the fastest
growing plane wave perturbation for a given $k_X$, $k_Z$ takes the
form
\begin{equation}
u = u_0\exp[ik_XX+ik_ZZ-\nu(k_X^2+k_Z^2)t],
\end{equation}
\begin{equation}
v = -2Atu_0\exp[ik_XX+ik_ZZ-\nu(k_X^2+k_Z^2)t],
\end{equation}
\begin{equation}
w = -{k_X\over k_Z}u_0\exp[ik_XX+ik_ZZ-\nu(k_X^2+k_Z^2)t],
\end{equation}
\begin{equation}
\hat{p} = 0,
\end{equation}
where $u_0$ is an arbitrary amplitude.  The ratio of energy at
time $t$ to the initial energy is then given by
\begin{equation}
G(t) = {u^2(t)+v^2(t)+w^2(t) \over u^2(0)+v^2(0)+w^2(0)} =
\left[1+{k_Z^2\over k_X^2+k_Z^2}(2At)^2\right]
\exp\left[-2\nu(k_X^2+k_Z^2)t\right].
\end{equation}
Since we are interested in flows that have perturbations with large
growth, let us ignore the 1 in the first factor.  Then, the time
at which the energy is maximum is given by
\begin{equation}
t_{max} = {1\over\nu\left(k_X^2+k_Z^2\right)},
\end{equation}
and the corresponding energy growth factor is
\begin{equation}
G_{max} = \left({2A\over\nu}\right)^2\,e^{-2}
{k_Z^2\over (k_X^2+k_Z^2)^3}.
\end{equation}

The problem we have analysed in \S3 is a flow with walls at $X=\pm L$
with no-slip boundary conditions.  In the absence of viscosity, the
simplest solution to this problem is
\begin{equation}
u = -{k_Z\over k_X}w = u_0\left(1+\cos{\pi X\over L}\right)\exp(ik_ZZ),
\qquad v=\hat{p}=0,
\end{equation}
which can be seen by inspection to satisfy the boundary conditions at
$X=\pm L$.  This solution is the sum of a plane wave with $k_X=0$ and
amplitude $u_0$, and two waves with $k_X=\pm \pi/L$ and amplitude
$u_0/2$.  Roughly, we expect that the solutions for $t_{max}$,
$G_{max}$ we wrote down earlier for a single plane wave will be
approximately correct provided we set $k_X^2$ equal to $(\pi/L)^2/2$,
i.e., the mean of 0 and $(\pi/L)^2$.  Noting that the Reynolds number
is given by
\begin{equation}
R = {|2A|L^2\over\nu},
\end{equation}
we find that the maximum growth factor is given by
\begin{equation}
G_{max}(k_Z,R) = {R^2k_Z^2L^2\,e^{-2}\over \left[{1\over2}\pi^2
+k_Z^2L^2\right]^3}.
\end{equation}
Maximizing this over $k_Z$, we find that the optimum wavevector is
\begin{equation}
k_ZL = k_z=\pi/2 = 1.57.
\end{equation}
This is close to the numerically determined value of 1.6 given in
Table 1.  The maximum growth factor and time at which this maximum is
attained are then
\begin{equation}
G_{max}(R) = 0.82\times10^{-3}R^2,
\end{equation}
\begin{equation}
|2A|t_{max}(R) = 0.13 R.
\end{equation}
These relations are shown as dashed lines in Figure \ref{fig3}.  We
see that the scaling with $R$ agrees well with the numerical
results, and the coefficient is also reasonably close.

\subsubsection{Constant Specific Angular Momentum Flow}

In this section we consider a rotating flow with $q=2$.  The vorticity
frequency $2B$ vanishes, and so the term proportional to it is not
present in equation (\ref{ymmtm}).  We can then write down the
following plane wave solution
\begin{equation}
u = {k_Z^2\over k_X^2+k_Z^2} 2\Omega t
v_0\exp[ik_XX+ik_ZZ-\nu(k_X^2+k_Z^2)t],
\end{equation}
\begin{equation}
v = v_0\exp[ik_XX+ik_ZZ-\nu(k_X^2+k_Z^2)t],
\end{equation}
\begin{equation}
w = -{k_Xk_Z\over k_X^2+k_Z^2} 2\Omega t
v_0\exp[ik_XX+ik_ZZ-\nu(k_X^2+k_Z^2)t],
\end{equation}
\begin{equation}
\hat{p} = -{ik_X\over k_X^2+k_Z^2} 2\Omega 
v_0\exp[ik_XX+ik_ZZ-\nu(k_X^2+k_Z^2)t],
\end{equation}
where $v_0$ is an arbitrary amplitude.

This solution looks quite different from that for plane Couette flow.
For instance, here $u$ and $w$ grow linearly with time at early time
and $v$ remains constant, which is the reverse of the case for plane
Couette flow.  Also, now we have a non-zero pressure perturbation.
Nevertheless, the energy growth factor has the same dependence on time
as for plane Couette flow:
\begin{equation}
G(t) = {u^2(t)+v^2(t)+w^2(t) \over u^2(0)+v^2(0)+w^2(0)} =
\left[1+{k_Z^2\over k_X^2+k_Z^2}(2\Omega t)^2\right]
\exp\left[-2\nu(k_X^2+k_Z^2)t\right].
\end{equation}
Note that, for $q=2$, $2\Omega=2A$, so the result is in fact
identical.  The reason for this close similarity is apparent when one
considers the original dynamical equations
(\ref{xmmtm})--(\ref{zmmtm}).  The only difference between plane
Couette flow and $q=2$ flow is that for the former the term $2\Omega
v$ is missing in equation (\ref{xmmtm}) and the term $-2Bu$ present
and is equal to $-2Au$ in equation (\ref{ymmtm}), whereas for the
latter the term $2\Omega v$ is present and is equal to $2Av$ in
equation (\ref{xmmtm}) and the term $-2Bu$ is missing in equation
(\ref{ymmtm}).  The equations are thus very symmetrical, except that
$X$ and $Y$ are interchanged in the two cases.  The resulting flows
look very different because of the switch in coordinates, but the
growth is identical.

The rest of the analysis proceeds exactly as in the previous
subsection.  As before, we conclude that the optimum $k_ZL=k_z \sim
1.57$, that the maximum growth factor is $G_{max}\sim
0.82\times10^{-3}R^2$, and that the maximum growth happens at a time
$t_{max}\sim 0.13R$.  As we saw in \S3, the numerical results are
indeed very similar for plane Couette flow and $q=2$ flow (Figure 3).  
The present analysis explains why this happens even though
the dynamics are quite different.

\subsubsection{$q<2$ Flow}

Now we consider a more general flow with $q<2$.  Such a flow has an
angular momentum gradient that is stable according to the Rayleigh
criterion.  This leads to epicyclic oscillations with frequency
$\kappa$ [see Eq. (\ref{epicyclic})].  

Let us ignore viscosity.  Then, defining $k\equiv
(k_X^2+k_Z^2)^{1/2}$, the plane wave solution with maximum growth is
given by
\begin{equation}
u = {k_Z\over k} {2\Omega\over \kappa} v_0 \exp(ik_XX+ik_ZZ)
\sin\left({k_Z\over k}\kappa t\right),
\end{equation}
\begin{equation}
v = v_0 \exp(ik_XX+ik_ZZ) \cos\left({k_Z\over k}\kappa t\right),
\end{equation}
\begin{equation}
w = -{k_X\over k} {2\Omega\over \kappa} v_0 \exp(ik_XX+ik_ZZ)
\sin\left({k_Z\over k}\kappa t\right),
\end{equation}
\begin{equation}
\hat{p} = -{ik_X\over k^2} 2\Omega v_0 \exp(ik_XX+ik_ZZ)
\cos\left({k_Z\over k}\kappa t\right).
\end{equation}
The energy growth as a function of time is given by
\begin{equation}
G(t) = \cos^2\left({k_Z\over k}\kappa t\right) +
{2\over (2-q)} \sin^2\left({k_Z\over k}\kappa t\right).
\end{equation}
Clearly, the maximum possible growth is
\begin{equation}
G_{max} = {2\over 2-q}, \label{epicyclicgrowth}
\end{equation}
and the growth happens after a time proportional to one quarter the
epicyclic period,
\begin{equation}
t_{max} = {\pi\over 2\kappa}\left(\frac{k}{k_Z}\right).
\end{equation}
In the actual flow with walls and viscosity the growth will be a
little less, as confirmed by the numerical results in Table 1, but
equation (\ref{epicyclicgrowth}) gives a rigorous upper limit to the
growth.

The key point of this analysis is that, for $q<2$, there is a limit to
the growth that arises just from dynamics; specifically, it is caused
by the presence of a non-vanishing epicyclic frequency.  Moreover, the
limiting growth in energy is just a factor of 4 for a Keplerian flow.
The limit has nothing to do with viscosity.  In contrast, both plane
Couette flow and $q=2$ flow can have infinite growth as far as the
dynamics are concerned and the limit to growth arises only from
viscosity.  

Balbus, Hawley \& Stone (1996) and Hawley, Balbus \& Winters (1999) suggested that the
existence of epicyclic motion lends dynamical stability to flows with
$q<2$ and that this makes these flows more resistant to turbulence.
Our analysis confirms their suggestion for perturbations with vertical
structure.  However, their argument does not apply to the
two-dimensional perturbations we consider next.

\subsection{Two-Dimensional Perturbations}

The last subsection explained why perturbations with vertical
structure have very little growth for $q<2$.  The growth is especially
insignificant for a Keplerian flow.  \S3 showed that these flows have
more growth for perturbations with non-zero $k_Y$.  In fact, for a
Keplerian flow, the maximum growth is for $k_Z=0$, $k_Y \ne 0$.  We
now consider such perturbations.

We consider a plane wave that is frozen into the fluid and is sheared
along with the background flow (see Figs. \ref{fig7}, \ref{fig8}).  If the flow starts
at time $t=0$ with initial wave-vector $(k_{Xi}, k_Y)$ in the
$XY$-plane, then the $X$-wavevector at later times is given by
\begin{equation}
k_X(t) = k_{Xi} + q\Omega k_Yt.
\label{kx0}
\end{equation}
With the above definition of $k_X$, we consider a plane wave solution
of the form
\begin{equation}
u = u(t) \exp(ik_XX+ik_YY), ~{\rm etc.}.
\end{equation}
Because of the non-zero $k_Y$, the Lagrangian time derivative is given
by
\begin{equation}
{d\over dt} = {\partial\over \partial t}-iq\Omega k_YX.
\end{equation}

The relevant plane wave solution in the absence of viscosity has been
written down by a number of authors (e.g., Chagelishvili et al. 2003;
Umurhan \& Regev 2004).  Generalizing the solution for finite
viscosity, we have
\begin{equation}
u = \zeta {k_Y\over k^2}
\exp\left(ik_XX+ik_YY-\nu\int_0^tk^2(t')dt'\right),
\end{equation}
\begin{equation}
v = -\zeta {k_X\over k^2}
\exp\left(ik_XX+ik_YY-\nu\int_0^tk^2(t')dt'\right),
\end{equation}
\begin{equation}
w = 0,
\end{equation}
\begin{equation}
\hat{p} = i\zeta \left({1\over k^2}2\Omega - {k_Y^2\over k^4} 2q\Omega\right)
\exp\left(ik_XX+ik_YY-\nu\int_0^tk^2(t')dt'\right),
\end{equation}
where $\zeta$ is the amplitude of the vorticity perturbation.
Since $w=0$, we see that the perturbations are two-dimensional (hence
the name ``two-dimensional perturbations'').  Also, the velocity
components are independent of $q$ and it is the pressure that adjusts
so as to keep the dynamics the same for all values of $q$.  In fact,
the above solution is valid even for plane Couette flow, provided we
make the replacements $2\Omega\to 0$ and $2q\Omega\to -2A$.

In the absence of viscosity, the energy growth is given by
\begin{equation}
G(t) = {k_{Xi}^2+k_Y^2 \over k_X^2+k_Y^2},
\end{equation}
that is, the energy is inversely proportional to the square of the
total wave-vector $k^2=k_X^2+k_Y^2$.  This result is easy to
understand.  For inviscid incompressible two-dimensional flow, the
vorticity $\nabla\times \vec \vartheta$ is exactly conserved.  This
means that $k\vartheta$ is constant, so the velocity scales inversely
as $k$.  The energy must then vary as $k^{-2}$.  The energy is thus
largest when $k$ is smallest.  Using equation (\ref{kx0}) we now see
what is required if we wish to obtain a large energy growth.  We need
to start with a large negative value for $k_{Xi}$.  As time goes on,
$k_X$ will become progressively less negative; as a result, $k$ will
decrease and $G$ will increase.  The maximum growth will be achieved
when $k_X=0$, giving
\begin{equation}
G_{max} = 1+{k_{Xi}^2\over k_Y^2}, \qquad
|2A|t_{max} = {k_{Xi}\over k_Y}.
\end{equation}

Now consider the effect of having rigid walls at $X=\pm L$.  By the
uncertainty principle, $k_X$ cannot become exactly zero, but must have
a minimum magnitude, $k_{X,min}\sim\pi/L$.  The maximum energy
growth is then approximately given by
\begin{equation}
G_{max}(k_{Xi},k_Y,R=0) = {k_{Xi}^2+k_Y^2 \over k_{X,min}^2+k_Y^2}
\sim {k_{Xi}^2 \over (\pi/L)^2+k_Y^2},
\end{equation}
where we have assumed that $k_{Xi}\gg k_Y$.  Including also
the effect of viscosity this becomes
\begin{equation}
G_{max}(k_{Xi},k_Y,R) \sim {k_{Xi}^2L^2 \over \pi^2+k_Y^2L^2}
\exp\left(-{2\over 3R}{k_{Xi}^3L^2\over k_Y}\right).
\end{equation}
Maximizing this with respect to $k_{Xi}$ and $k_Y$, we obtain
\begin{equation}
k_YL=k_y \sim {\pi\over \sqrt{2}} = 2.2,
\qquad k_{Xi}L =k_{xi}\sim 1.3 R^{1/3}.
\end{equation}
The maximum growth and the time of maximum are then
\begin{equation}
G_{max}(R) \sim 0.059R^{2/3}, \qquad |2A|t_{max}(R) \sim 0.59 R^{1/3}.
\end{equation}
While the scalings with $R$ are accurate and agree with the numerical
results presented in \S3, the coefficients are approximate since they
depend on the assumed value of $k_{X,min}$.  If instead of taking
$k_{X,min}L=\pi$, we select $k_{X,min}L=1$, we find $k_YL =0.71$,
$k_{Xi}L = 0.89 R^{1/3}$, $G_{max}(R) \sim 0.27R^{2/3}$,
$|2A|t_{max}(R) \sim 1.25R^{1/3}$.  The results in \S3 lie between
these two estimates.  In fact, if we choose $k_{X,min}L\sim1.7$, then we
obtain $k_YL=1.2$, $G_{max}(R)\sim 0.13R^{2/3}$, $|2A|t_{max}(R)\sim 0.88R^{1/3}$,
which agree with the numerical result (Table 2).

Finally, we can carry out the analysis by keeping $k_Y$ fixed and
optimizing only $k_{Xi}$.  We then find that $G_{max}(k_Y,R)$ and
$t_{max}(k_Y,R)$ vary as
\begin{equation}
G_{max}(k_Y,R) = {(k_YL)^{2/3} \over (k_{X,min}L)^2+(k_YL)^2}
\exp\left(-{2\over3}\right)R^{2/3},
\end{equation}
\begin{equation}
|2A|t_{max}(k_Y,R) = (k_YL)^{-2/3} R^{1/3}.
\end{equation}
We see that $G_{max}$ varies as $(k_YL)^{2/3}$ for small $k_YL$ and as
$(k_YL)^{-4/3}$ for large $k_YL$.  The time of maximum scales as
$(k_YL)^{-2/3}$ for all $k_YL$.  The above analytical results are
plotted as dotted lines in Figure \ref{fig6}, assuming
$k_{X,min}L=1.7$ as derived above.  We see that the agreement with the
numerical results is very good except at very large $R$ where the
calculations are not very accurate.

\section{Discussion and Conclusions}

We have demonstrated that significant transient growth of
perturbations is possible in a Keplerian flow between walls (as shown
by Yecko 2004). Although the system does not have any unstable
eigenmodes, nevertheless, because of the non-normal nature of the
eigenmodes a significant level of transient energy growth is possible
for appropriate choice of initial conditions. If the maximum growth
exceeds the threshold for inducing turbulence, it is plausible that
this mechanism could drive the system to a turbulent state.
Presumably, once the system becomes turbulent it can remain turbulent
as a result of nonlinear interactions and feedback among the
perturbations.

In this so-called bypass mechanism for transition to turbulence, the
maximum energy growth and the time needed for this growth are likely to be
the main factors that control the transition to hydrodynamic
turbulence.  It has been observed in laboratory experiments that plane
Couette flow can be made turbulent for Reynolds numbers above a
critical value $R_c\sim 350$.  According to our analysis, for $R=350$,
the maximum energy growth is $G_{max}(R=350)=145$, and the maximum
occurs at time $t_{max}=42.3$ (Fig. 3).  Since a constant
angular momentum disk ($q=2$) behaves very similarly to plane Couette
flow, the critical Reynolds number for turbulence for this case is
also likely to be $R_c\sim 350$.  For this $R$, the growth factor is
$G_{max}(R=350)=143.5$ and the time-scale is $t_{max}=48.3$.

It is true that the underlying equations as well as the maximally
growing modes in plane Couette flow (and a q=2 disk) are different
compared to a Keplerian flow, and so the turbulent phases in the two
systems may have significant differences.  However, the presence of
similar boundary conditions may suggest similarities in the kinematic
structure of the turbulence. In fact, a Keplerian disk and a constant
angular momentum disk are two special cases of rotating shearing flows
parameterized by $q$. Based on this, we make the plausible assumption
that the threshold energy growth factor needed for transition to
turbulence in a shear flow with any value of $q$ is $E_c \sim 145$.
Applying this conjecture to the optimal two-dimensional perturbations
of a Keplerian disk analysed in \S 4.2, we estimate the critical
Reynolds number for a Keplerian flow to be $R_c\sim 3.4\times10^4$,
i.e., a factor of 100 greater than for plane Couette flow.  The time
to reach the maximum is $t_{max}=28.3$, which is comparable to
that in plane Couette flow, and is not too large compared to the
accretion time scale of a geometrically thin disk.

Instead of taking $R_c\sim350$, which is perhaps somewhat optimistic
since plane Couette flow needs to be perturbed significantly before it
will become turbulent at this Reynolds number, we might wish to be
conservative and assume $R_c\sim1000$ for this flow.  At this value of
$R$, plane Couette flow and $q=2$ flow have $G_{max}(R=1000)\sim 1200$
and $t_{max}\sim120-140$.  Applying the requirement $E_c\sim1200$ to
Keplerian flow, we find $R_c\sim 10^6$ and $t_{max}\sim 100$.  Now the
critical Reynolds number is a factor of 1000 greater than for plane
Couette flow.

Why is the critical Reynolds number so much larger for
a Keplerian disk compared to a constant angular momentum disk or plane
Couette flow?  The numerical results in \S 3 and the analytical work
in \S 4 provide the answer, viz., the presence of epicyclic motions in
a Keplerian disk.  It is very interesting to note that the presence of
epicyclic motion not only kills growth dramatically, it also changes
the optimum wavevector $\{k_y,k_z\}$ of the perturbations needed to
produce energy growth.  For a constant angular momentum disk ($q=2$)
and plane Couette flow, both of which have zero epicyclic frequency,
it is seen that growth is maximized for $k_y\sim0$ (on the $k_z$
axis).  Even for a very small shift in the value of $q$ below 2,
corresponding to the introduction of a small epicyclic frequency, the
location of maximum growth immediately moves significantly in the
$k_y-k_z$ plane from the $k_z$ axis (see Fig. 4 which corresponds to
$R=2000$). With decreasing $q$, the epicyclic motion of the disk
increases, and correspondingly the optimum value of $k_y$ for growth
increases while the optimum $k_z$ decreases. When $q=1.5$, i.e., when
the disk is exactly Keplerian, the growth is maximum for $k_z=0$ (on
the $k_y$ axis). To the best of our knowledge, this change in the
location of the maximum in the $k_y-k_z$ plane has not been commented
upon prior to this work.

The change between $q=2$ and $q=1.5$ may be completely understood
analytically, as we show in \S 4.  The important point is that the
vertical perturbations ($k_y=0$) that cause the large observed growth
in a $q=2$ disk require an absence of epicyclic motions.  When the
epicyclic frequency is zero, the velocity perturbation is able to grow
linearly with time and the energy grows quadratically.  The only limit
to growth is provided by viscosity, which gives a scaling
$G_{max}\propto R^2$.  However, once there is a non-zero epicyclic
frequency, the growth is immediately limited.  Even in the absence of
viscosity, only a modest level of growth is possible.  In fact, for a
Keplerian flow, the maximum growth that one can obtain from vertical
perturbations is only 4, well below the critical growth needed for
turbulence.  If vertical perturbations were the sole route to
turbulence, then a Keplerian flow could never make the transition to
turbulence.

However, as \S 4.2 shows, there are other kinds of perturbations,
specifically two-dimensional perturbations with $k_z=0$, which are not
affected by epicyclic motions.  For these perturbations, pressure
fluctuations are able to absorb the effect of the Coriolis force.  As
a result, two-dimensional perturbations are able to grow to
arbitrarily large values in the absence of viscosity.  However, the
growth is much reduced compared to the vertical perturbations
described in the previous paragraph and it scales only as $R^{2/3}$.
Thus, one needs much larger values of $R\sim10^{4.5}-10^6$ before one
can achieve the same level of energy growth as can be found in a $q=2$
disk for Reynolds numbers as small as $10^{2.5}-10^3$.

These results lead to a better understanding of the numerical
simulations described in Balbus, Hawley \& Stone (1996) and Hawley, Balbus
\& Winters (1999).  Both papers showed that there is a close similarity
between plane Couette flow and $q=2$ flow, in the sense that the two
flows readily became turbulent in numerical simulations.  However,
once the authors reduced the value of $q$ below about 1.95, no
turbulence was seen even when the flows were initialized with large
perturbations.  The authors suggested that the change in behavior is
because of the dynamical stability imposed by the Coriolis force and
epicyclic motions.  Our analysis supports this conclusion.

However, Balbus, Hawley \& Stone (1996) and Hawley, Balbus \& Winters 
(1999) then proceeded
to rule out the possibility of hydrodynamic turbulence in Keplerian
disks.  We do not agree with this conclusion.  As we have shown,
Keplerian disks can indeed support large transient energy growth, but
they need much larger Reynolds numbers to achieve the same energy
growth as plane Couette flow or $q=2$ flow.  The numerical simulations
probably had effective Reynolds numbers $\lesssim10^4$ (because of
numerical viscosity) which is below our most optimistic estimate of
the critical Reynolds number.  Thus, we suspect the simulations 
did not have sufficient numerical resolution to permit turbulence.
In fact, Longaretti (2002) already suspected that the non occurrence of
turbulence in previous simulations may be just due to the choice
of low Reynolds number.

Although the problem we analysed is shear flow between walls, the
optimum growing perturbations that we find for the Keplerian case are
very similar to those described by Chagelishvili et al. (2003) and
Umurhan \& Regev (2004) for an infinite shear flow.  The perturbations
are basically plane waves that are frozen in the shearing flow.
Initially, at $t=0$, the effective wave vector of the perturbation in
the $x$ direction ($k_x$) is negative, which means that we have very
asymmetric leading waves.  As time goes on, the wavefronts are
straightened out by the shear and $|k_x|$ decreases. At the time when
the growth is maximum, $k_x\sim 0$ (but not precisely 0 because of the
walls, see \S 4.2) and the wavefronts become almost radial.  At yet
later time, the growth decreases and the wave becomes of a trailing
pattern.

The above time evolution is very different from that seen for the
optimum perturbations in plane Couette flow or in a $q=2$ disk.  In
plane Couette flow, the $x$-component of the perturbation, $u$
(i.e. the normal velocity), dominates over the other components, $v$,
$w$, at $t=0$.  However, $u$ remains at the same level for all time
whereas $v$ and $w$ increase strongly up to the point of maximum
growth before declining.  The overall shape of the perturbation is
roughly self-similar with time.  For a constant angular momentum disk,
on the other hand, it is $v$ which remains constant with time whereas
$u$ and $w$ vary by large amounts.  However, as in plane Couette flow,
the solution is largely self-similar in character up to the maximum.
Neither of these flows shows the shearing perturbations that are
characteristic of the Keplerian problem (Figs. 7 and 8).

We conclude with an important caveat.  While the demonstration of
large energy growth is an important step, it does not prove that
Keplerian disks will necessarily become hydrodynamically turbulent.
Umurhan \& Regev (2004) have shown via two-dimensional simulations
that chaotic motions can persist for a time much longer than the time
scale $t_{max}$ needed for linear growth.  However, they also note
that their perturbations must ultimately decline to zero in the
presence of viscosity.  To overcome this limitation, it is necessary
to invoke three-dimensional effects.  Secondary instabilities of
various kinds, such as the elliptical instability, are widely
discussed as a possible route to self-sustained turbulence in linearly
perturbed shear flows (see the review article by Kerswell (2002); see
also e.g. Hellberg \& Orszag 1988; Le Dize\'s \& Rossi 1996).  It
remains to be seen if these three-dimensional instabilities are
present in perturbed flows such as those shown in Figures 7 and 8.  If
they are, one will in addition have to show that they lead to
non-linear feedback and 3-dimensional turbulence.

\begin{acknowledgements}
We would like to thank the referee for various suggestions that
improved the presentation of the paper.
This work was supported in part by NASA grant NAG5-10780 and NSF grant
AST 0307433.

\end{acknowledgements}

\appendix
\section{Appendix: Method to Compute the Transient Growth}

To compute the optimum growth, first we need to evaluate the 2-norm of $Q$.
From (\ref{grow1}) it is clear that the 2-norm depends on ${\cal L}$ which consists
of ${\cal L}_{os}$ and ${\cal L}_{sq}$. The underlying Hilbert space
of the second order linear operator, ${\cal L}_{sq}$, is ${\cal H}_{sq}
=L^2[-1,1]$\,\footnote{The Hilbert space is defined as a complete vector space
with an inner product.}. Therefore the inner product of
$\hat{\zeta}_1,\hat{\zeta}_2\in {\cal H}_{sq}$ is defined as
\begin{eqnarray}
(\hat{\zeta}_1,\hat{\zeta}_2)_L=\int_{-1}^{1}\hat{\zeta}_2^*\hat{\zeta}_1dx.
\label{sqin}
\end{eqnarray}
The domain of ${\cal L}_{sq}$, that is ${\cal D}_{sq}$, is the set of functions
$\{\psi\}$ which have a second derivative in $L^2[-1,1]$ satisfying $\psi(\pm 1)=0$.
Following DiPrima \& Habetler (1969) we also define the underlying
Hilbert space of ${\cal L}_{os}$, ${\cal H}_{os}$, consisting of the set of
functions $\{\psi\}$ having a second derivative in $L^2[-1,1]$ satisfying
$\psi(\pm 1)=0$. Therefore, for $\hat{u}_1,\hat{u}_2\in {\cal H}_{os}$ the
inner product is defined as
\begin{eqnarray}
(\hat{u}_1,\hat{u}_2)_H=(D\hat{u}_1,D\hat{u}_2)_L+k^2(\hat{u}_1,\hat{u}_2)_L.
\label{osin}
\end{eqnarray}
The domain of ${\cal L}_{os}$, that is ${\cal D}_{os}$, is the set of functions
$\{\psi\}$ that have a fourth derivative in $L^2[-1,1]$ satisfying $\psi(\pm 1)=
\psi^\prime(\pm 1)=0$. Therefore the underlying Hilbert space of $\cal L$ is
${\cal H}={\cal H}_{os}\times{\cal H}_{sq}$ and the corresponding domain is
${\cal D}={\cal D}_{os}\times {\cal D}_{sq}$. Thus combining (\ref{sqin})
and (\ref{osin}) and with some algebra, the inner product for $Q_1,Q_2\in{\cal H}$
can be written as
\begin{eqnarray}
(Q_1,Q_2)=({\cal F}\hat{u}_1,\hat{u}_2)_L+(\hat{\zeta}_1,\hat{\zeta}_2)_L,
\label{inner}
\end{eqnarray}
where ${\cal F}=-(D^2-k^2)$.

Now following Butler \& Farrell (1992) the perturbation energy density can be evaluated as
\begin{eqnarray}
E=\frac{1}{2V}\int_{-1}^{1}\int_0^a\int_0^b (u^2+v^2+w^2)dzdydx,
\label{peren}
\end{eqnarray}
where $a=2\pi/k_y$, $b=2\pi/k_z$, and $V=2ab$ is the integration volume.
The physical velocity components are the real quantity obtained as
\begin{eqnarray}
u=\frac{1}{2}\{\hat{u}\,\exp[i{\vec k}.{\vec r}_p]+\hat{u}^*\exp[-i{\vec k}.{\vec r}_p]\}.
\label{ureal}
\end{eqnarray}
Now replacing $(v^2+w^2)$ in terms of $\zeta$ and $du/dx$ in (\ref{peren})
and integrating over $y$ and $z$ we obtain
\begin{eqnarray}
E=\frac{1}{8k^2}\int_{-1}^{1}\left[k^2\hat{u}^\dag \hat{u}+\frac{\partial
\hat{u}}{\partial x}^\dag\frac{\partial \hat{u}}{\partial x}+\hat{\zeta}^\dag\hat{\zeta}\right]dx,
\label{peren2}
\end{eqnarray}
where $\hat{u}$ and $\hat{\zeta}$ are considered to be $N$ dimensional column matrices.
Now combining (\ref{sol8}), (\ref{inner}) and (\ref{peren2}) we obtain
\begin{eqnarray}
8k^2E=||Q_K||^2=C^\dag e^{i\Sigma_K t}\hat{Q}e^{-i\Sigma_K t}C,
\label{peren3}
\end{eqnarray}
where $\hat{Q}$ is a $K\times K$ Hermitian matrix whose $ij$th element is the inner
product of $\tilde{Q}_i$ and $\tilde{Q}_j$,
\begin{eqnarray}
\hat{Q}_{ij}=(\tilde{Q}_i,\tilde{Q}_j)=({\cal F}\tilde{u}_i,\tilde{u}_j)_L+(\tilde{\zeta}_i,\tilde{\zeta}_j)_L.
\label{inij}
\end{eqnarray}
Decomposing $\hat{Q}$ in terms of a matrix $W$ according to
$\hat{Q}=W^\dag W$, and combining (\ref{grow1}) and (\ref{peren3}) ,
we obtain the expression for the optimum growth (see also Reddy \& Henningson 1993 and
Schmid \& Henningson 1994),
\begin{eqnarray}
G_K(t)={\rm maximum}\left(\frac{||W\exp[-i\Sigma_K t] C||^2_2}{||WC||^2_2}\right)
=||W\,\exp[-i\Sigma_K t]\,W^{-1}||^2_2,
\label{grow3}
\end{eqnarray}
where the subscript $2$ denotes the 2-norm or Euclidian norm. The 2-norm of the
matrix $W\,\exp[-i\Sigma_K i]\,W^{-1}$ can be evaluated by means of a singular value
decomposition. Then, for a given $t$, the square of the highest singular value is the maximum
energy growth, $G_{max}(t)$, for that time.
$W$ can be computed easily by a similarity
transformation of $\hat{Q}$
\begin{eqnarray}
\hat{Q}=S\sqrt{\hat{Q}_d}S^\dag S\sqrt{\hat{Q}_d}S^\dag=W^\dag W,
\label{sim}
\end{eqnarray}
where $S$ is a unitary matrix and
$\hat{Q}_d$ is a diagonal matrix consisting of the eigenvalues of $\hat{Q}$
along the diagonal. Therefore one only needs to construct the matrix $\hat{Q}$
to compute $G_K(t)$, while $\Sigma_K$ is immediately available from the
eigenvalues of ${\cal L}$. From (\ref{peren2}), (\ref{peren3}) and (\ref{inij}),
in the finite-difference approximation $\hat{Q}$ can be written as
\begin{eqnarray}
\nonumber
\hat{Q}&=&\Delta x\left[k^2 U^\dag U+\frac{\partial U}{\partial x}^\dag\frac{\partial U}
{\partial x}+Z^\dag Z\right],\\
U_{mj}&=&\tilde{u}_{mj};\hskip0.5cm Z_{mj}=\tilde{\zeta}_{mj}
\label{ee}
\end{eqnarray}
where $m$ runs between $1$ and $N$ (points on the finite-difference grid)
and $j$ ranges from $1$ to $K$ (mode numbers).

{}


\begin{thebibliography}{}

\bibitem[]{} Afshordi, N., Mukhopadhyay, B. \& Narayan, R. 2005, ApJ (submitted); AMN05.
\bibitem[]{} Balbus, S. \& Hawley, J. 1991, ApJ, 376, 214.
\bibitem[]{} Balbus, S., Hawley, J., \& Stone, J. 1996, ApJ, 467, 76.
\bibitem[]{} Bech, K. \& Andersson, H. 1997, J. Fluid Mech., 347, 289.
\bibitem[]{} Blaes, O. \& Balbus, S. 1994, ApJ, 421, 163.
\bibitem[]{} Butler, K. \& Farrell, B. 1992, Phys. Fluids A, 4(8), 1637.
\bibitem[]{} Chagelishvili, G., Zahn, J.-P., Tevzadze, A. \&
Lominadze, J. 2003, A\&A, 402, 401.
\bibitem[]{} Chandrasekhar, S. 1960, Proc. Natl. Acad. Sci., 46, 53.
\bibitem[]{} DiPrima, R. C., \& Habetler, G. J. 1969, Arch. Rat. Mech. Anal., 34, 218.
\bibitem[]{} Drazin, P. \& Reid, W. 1981, in {\it Hydrodynamic Stability}, Cambridge.
\bibitem[]{} Farrell, B. 1988, Phys. Fluids, 31, 2093.
\bibitem[]{} Fleming, T. P., Stone, J. M., \& Hawley, J. F. 2000, ApJ, 530, 464
\bibitem[]{} Fromang, S., Terquem, C. \& Balbus, S. 2002, MNRAS, 329, 18.
\bibitem[]{} Gammie, C. 1996, ApJ, 457, 355.
\bibitem[]{} Gammie, C. \& Menou, K. 1998, ApJ, 492, L75.
\bibitem[]{} Goldreich, P. \& Lynden-Bell, D. 1965, MNRAS, 130, 125.
\bibitem[]{} Goldreich, P. \& Tremaine, S. 1978, ApJ, 222, 850; 1979, ApJ, 233, 857.
\bibitem[]{} Goodman, J. 2003, MNRAS, 339, 937.
\bibitem[]{} Hawley, J., Balbus, S. \& Winters, W. 1999, ApJ, 518, 394.
\bibitem[]{} Hawley, J., Gammie, C. \& Balbus, S. 1995, ApJ, 440, 742.
\bibitem[]{} Hawley, J., Gammie, C. \& Balbus, S. 1996, ApJ, 464, 690.
\bibitem[]{} Hellberg, C. \& Orszag, S. 1988, Phys. Fluids, 31(1), 6. 
\bibitem[]{} Ioannaou, P. \& Kakouris, A. 2001, ApJ, 550, 931.
\bibitem[]{} Johnson, B., \& Gammie, C. 2005a, ApJ (submitted), astro-ph/0501005;
2005b, ApJ (submitted). 
\bibitem[]{} Landau, L., \& Lifshitz, E. 1989, in {\it Fluid Mechanics, Second Edition},
Oxford: Butterworth-Heinemann.
\bibitem[]{} Le Dize\'s, S., Rossi, M. \& Moffatt, K. 1996, Phys. Fluids, 8(8), 2084.
\bibitem[]{} Longaretti, P. 2002, ApJ, 576, 587.
\bibitem[]{} Lynden-Bell, D. \& Pringle, J. 1974, MNRAS, 168, 603.
\bibitem[]{} Kerswell, R. 2002, Ann. Rev. Fluid Mech., 34, 83.
\bibitem[]{} Menou, K. 2000, Science, 288 (5473), 2022.
\bibitem[]{} Menou, K., \& Quataert, E. 2001, ApJ, 552, 204.
\bibitem[]{} Narayan, R., Goldreich, P., \& Goodman, J. 1987, MNRAS, 228, 1.
\bibitem[]{} Orr, W. 1907, Proc. R. Irish. Acad. A, 27, 9.
\bibitem[]{} Orszag, S. 1971, J. Fluid Mech., 50, 689.
\bibitem[]{} Reddy, S. \& Henningson, D. 1993, J. Fluids Mech., 252, 209.
\bibitem[]{} Richard, D. \& Zahn, J.-P. 1999, A\&A, 347, 734.
\bibitem[]{} Romanov, V. 1973, Funct. Anal. Appl., 7, 137.
\bibitem[]{} Schmid, P., \& Henningson, D. 1994, J. Fluids Mech., 277, 197.
\bibitem[]{} Shakura, N., \& Sunyaev, R. 1973, A\&A, 24, 337.
\bibitem[]{} Swinney, H. \& Gollub, J. 1981, in {\it Hydrodynamic Instabilities
and the Transition to Turbulence}, Springer-Verlag.
\bibitem[]{} Tevzadze, A., Chagelishvili, G., Zahn, J.-P., Chanishvili, R., 
\& Lominadze, J. 2003, A\&A, 407, 779.
\bibitem[]{} Trefethen, L., Trefethen, A., Reddy, S. \& Driscoll, T. 1993,
Science, 261, 578.
\bibitem[]{} Umurhan, O., \& Regev, O. 2004, A\&A, 427, 855.
\bibitem[]{} Velikhov, E. 1959, J. Exp. Theor. Phys. (USSR), 36, 1398.
\bibitem[]{} Yecko, P. 2004, A\&A, 425, 385.

\end{thebibliography}
\end{document}